\documentclass[submission,copyright,creativecommons]{eptcs}
\providecommand{\event}{MSFP 2018} 
\usepackage{breakurl}              
\usepackage{pig}
\usepackage{stmaryrd}
\usepackage{upgreek}
\usepackage[all]{xy}

\ColourEpigram
\newcommand{\D}[1]{\blue{\ensuremath{\mathsf{#1}}}}
\newcommand{\C}[1]{\red{\ensuremath{\mathsf{#1}}}}
\newcommand{\F}[1]{\green{\ensuremath{\mathsf{#1}}}}
\newcommand{\V}[1]{\purple{\ensuremath{\mathit{#1}}}}

\newcommand{\OPE}[1]{\Updelta_{\!+}^{#1}}

\title{Everybody's Got To Be Somewhere}
\author{
  Conor McBride
  \institute{Mathematically Structured Programming Group\\
             Department of Computer and Information Sciences\\
             University of Strathclyde, Glasgow}
  \email{conor.mcbride@strath.ac.uk}
}
\def\titlerunning{Everybody's Got To Be Somewhere}
\def\authorrunning{C.T. McBride}
\makeatletter
\@ifundefined{lhs2tex.lhs2tex.sty.read}%
  {\@namedef{lhs2tex.lhs2tex.sty.read}{}%
   \newcommand\SkipToFmtEnd{}%
   \newcommand\EndFmtInput{}%
   \long\def\SkipToFmtEnd#1\EndFmtInput{}%
  }\SkipToFmtEnd

\newcommand\ReadOnlyOnce[1]{\@ifundefined{#1}{\@namedef{#1}{}}\SkipToFmtEnd}
\usepackage{amstext}
\usepackage{amssymb}
\usepackage{stmaryrd}
\DeclareFontFamily{OT1}{cmtex}{}
\DeclareFontShape{OT1}{cmtex}{m}{n}
  {<5><6><7><8>cmtex8
   <9>cmtex9
   <10><10.95><12><14.4><17.28><20.74><24.88>cmtex10}{}
\DeclareFontShape{OT1}{cmtex}{m}{it}
  {<-> ssub * cmtt/m/it}{}

\DeclareFontShape{OT1}{cmtt}{bx}{n}
  {<5><6><7><8>cmtt8
   <9>cmbtt9
   <10><10.95><12><14.4><17.28><20.74><24.88>cmbtt10}{}
\DeclareFontShape{OT1}{cmtex}{bx}{n}
  {<-> ssub * cmtt/bx/n}{}

\newcommand{\anonymous}{\kern0.06em \vbox{\hrule\@width.5em}}

\usepackage{polytable}

\@ifundefined{mathindent}%
  {\newdimen\mathindent\mathindent\leftmargini}%
  {}%

\def\resethooks{%
  \global\let\SaveRestoreHook\empty
  \global\let\ColumnHook\empty}
\newcommand*{\savecolumns}[1][default]%
  {\g@addto@macro\SaveRestoreHook{\savecolumns[#1]}}
\newcommand*{\restorecolumns}[1][default]%
  {\g@addto@macro\SaveRestoreHook{\restorecolumns[#1]}}
\newcommand*{\aligncolumn}[2]%
  {\g@addto@macro\ColumnHook{\column{#1}{#2}}}

\resethooks

\newcommand{\onelinecommentchars}{\quad-{}- }
\newcommand{\commentbeginchars}{\enskip\{-}
\newcommand{\commentendchars}{-\}\enskip}

\newcommand{\visiblecomments}{%
  \let\onelinecomment=\onelinecommentchars
  \let\commentbegin=\commentbeginchars
  \let\commentend=\commentendchars}

\newcommand{\invisiblecomments}{%
  \let\onelinecomment=\empty
  \let\commentbegin=\empty
  \let\commentend=\empty}

\visiblecomments

\newlength{\blanklineskip}
\newcommand{\hsindent}[1]{\quad}
\let\hspre\empty
\let\hspost\empty

\EndFmtInput
\makeatother
\ReadOnlyOnce{polycode.fmt}%
\makeatletter

\newcommand{\hsnewpar}[1]%
  {{\parskip=0pt\parindent=0pt\par\vskip #1\noindent}}

\newcommand{\hscodestyle}{}

\newcommand{\sethscode}[1]%
  {\expandafter\let\expandafter\hscode\csname #1\endcsname
   \expandafter\let\expandafter\endhscode\csname end#1\endcsname}

  {\par\noindent
   \advance\leftskip\mathindent
   \hscodestyle
   \let\\=\@normalcr
   \let\hspre\(\let\hspost\)%
   \pboxed}%
  {\endpboxed\)%
   \par\noindent
   \ignorespacesafterend}

  {\hsnewpar\abovedisplayskip
   \advance\leftskip\mathindent
   \hscodestyle
   \let\hspre\(\let\hspost\)%
   \pboxed}%
  {\endpboxed%
   \hsnewpar\belowdisplayskip
   \ignorespacesafterend}

  {\hsnewpar\abovedisplayskip
   \advance\leftskip\mathindent
   \hscodestyle
   \let\\=\@normalcr
   \(\pboxed}%
  {\endpboxed\)%
   \hsnewpar\belowdisplayskip
   \ignorespacesafterend}

\newcommand{\plainhs}{\sethscode{plainhscode}}

\plainhs

  {\hsnewpar\abovedisplayskip
   \advance\leftskip\mathindent
   \hscodestyle
   \let\\=\@normalcr
   \(\parray}%
  {\endparray\)%
   \hsnewpar\belowdisplayskip
   \ignorespacesafterend}

  {\parray}{\endparray}

  {\(\parray}{\endparray\)}

\def\codeframewidth{\arrayrulewidth}
\RequirePackage{calc}

  {\parskip=\abovedisplayskip\par\noindent
   \hscodestyle
   \arrayrulewidth=\codeframewidth
   \tabular{@{}|p{\linewidth-2\arraycolsep-2\arrayrulewidth-2pt}|@{}}%
   \hline\framedhslinecorrect\\{-1.5ex}%
   \let\endoflinesave=\\
   \let\\=\@normalcr
   \(\pboxed}%
  {\endpboxed\)%
   \framedhslinecorrect\endoflinesave{.5ex}\hline
   \endtabular
   \parskip=\belowdisplayskip\par\noindent
   \ignorespacesafterend}

\newcommand{\framedhslinecorrect}[2]%
  {#1[#2]}

  {\(\def\column##1##2{}%
   \let\>\undefined\let\<\undefined\let\\\undefined
   \newcommand\>[1][]{}\newcommand\<[1][]{}\newcommand\\[1][]{}%
   \def\fromto##1##2##3{##3}%
   }{\) }%

  {\let\orighscode=\hscode
   \let\origendhscode=\endhscode
   \def\endhscode{\def\hscode{\endgroup\def\@currenvir{hscode}\\}\begingroup}
   \orighscode\def\hscode{\endgroup\def\@currenvir{hscode}}}%
  {\origendhscode
   \global\let\hscode=\orighscode
   \global\let\endhscode=\origendhscode}%

\makeatother
\EndFmtInput
\DeclareMathAlphabet{\mathkw}{OT1}{cmss}{bx}{n}

\begin{document}
\maketitle

\begin{abstract}
The key to any nameless representation of syntax is how it indicates the variables we choose
to use and thus, implicitly, those we discard. Standard de Bruijn representations delay 
discarding \emph{maximally} till the \emph{leaves} of terms where
one is chosen from the variables in scope at the expense of the rest. Consequently, introducing new but unused
variables requires term traversal. This paper introduces a
nameless `\emph{co}-de-Bruijn' representation which makes the opposite canonical choice, delaying
discarding \emph{minimally}, as near as possible to the \emph{root}.  It is literate Agda: dependent types
make it a practical joy to express and be driven by strong intrinsic invariants
which ensure that scope is aggressively whittled down to just the \emph{support} of each subterm, in which every
remaining variable occurs somewhere. The construction is generic, delivering a
\emph{universe} of syntaxes with higher-order \emph{meta}variables, for which the appropriate notion of
substitution is \emph{hereditary}. The implementation of simultaneous substitution exploits tight scope control
to avoid busywork and shift terms without traversal. Surprisingly, it is
also intrinsically terminating, by structural recursion alone.
\end{abstract}


When I was sixteen and too clever by half, I wrote a text editor which cached
a plethora of useful but redundant pointers into the buffer, just to shave a handful
of nanoseconds off redisplay. Accurately updating these pointers at each
keystroke was a challenge which taught me the hard way about the value of simplicity.
Now, I am a dependently typed programmer. I do not keep invariants: invariants keep me.

This paper is about scope invariants in nameless representations of syntax. One motivation
for such is eliminating redundant name choice to make $\alpha$-equivalence trivial. Classic de Bruijn
syntaxes~\cite{deBruijn:dummies} replace name by number:
variable uses count either out from use to binding (\emph{indices}), or
in from root to binding (\emph{levels}). Uses are found at the
leaves of syntax trees, so any operation which modifies the sequence of variables
in scope requires traversal. E.g., consider this
$\beta$-reduction (under $\lambda x$) in untyped $\lambda$-calculus.
\newcommand{\fb}[1]{\framebox{\ensuremath{#1}}}
\[\begin{array}{rrcl}
\mbox{name}    & \lambda x.\: (\lambda y.\:\, y\:x\:(\lambda z.\: z\:(y\:z)))\:\underline{(x\:(\lambda v.\:v))}
               & \leadsto_\beta
               & \lambda x.\: \underline{(x\:(\lambda v.\:v))}\:x\:(\lambda z.\: z\:(\underline{(x\:(\lambda v.\:v))}\:z)\\
\mbox{index}   & \lambda \;\,.\: (\lambda \;\,.\: 0\:1\:(\lambda \;.\: 0\:(1\:0)))\:\underline{(0\:(\lambda \;.\:0))}
               & \leadsto_\beta
               & \lambda \;\,.\: \underline{(0\:(\lambda \;.\:0))}\:0\:(\lambda \;\,.\: 0\:(\underline{(1\:(\lambda \;\,.\:0))}\:0)\\
\mbox{level}   & \lambda \;\,.\: (\lambda \;\,.\: 1\:0\:(\lambda \;.\: 2\:(1\:2)))\:\underline{(0\:(\lambda \;.\:1))}
               & \leadsto_\beta
               & \lambda \;\,.\: \underline{(0\:(\lambda \;.\:1))}\:0\:(\lambda \;\,.\: 1\:(\underline{(0\:(\lambda \;\,.\:2))}\:1)\\
\end{array}\]
Underlining shows the movement of the substituted term. In the \emph{index} representation,
the free $x$ must be shifted when it goes under the $\lambda z$. With \emph{levels},
the free $x$ stays $0$, but the bound $v$ must be shifted under $\lambda z$,
and the substitution context must be shifted to account for the eliminated $\lambda y$.
Shift happens.

The objective of this paper is not to eliminate shifts altogether, but to ensure that they
do not require traversal. The approach is to track exactly which variables are \emph{relevant}
at all nodes in the tree and aggressively expel those unused in any given subtree. As we do
so, we need and obtain much richer accountancy of variable usage, with much more intricate
invariants. Category theory guides the design of these invariants and Agda's dependent
types~\cite{DBLP:conf/afp/Norell08} drive their correct implementation.

\newcommand\bs{\char`\\}
My explorations follow Sato, Pollack, Schwichtenberg and
Sakurai, whose $\lambda$-terms make
binding sites carry \emph{maps} of use sites~\cite{spss:lambda.maps}. E.g.,
the \ensuremath{\F{\mathbb{K}}} and \ensuremath{\F{\mathbb{S}}} combinators become (respectively)
\[\begin{array}{r@{\qquad}r@{\:}r@{\:}l@{\qquad}r@{\:}r@{\:}r@{\:}c@{\:}c}
\mbox{names} & \lambda c.&\lambda e.&c & \lambda f.&\lambda s.&\lambda e.&(f\:e)&(s\:e)\\
\mbox{maps} & \texttt{1\bs}&\texttt{0\bs}&\square & \texttt{((10) (00))\bs}&\texttt{((00) (10))\bs}&\texttt{((01) (01))\bs}&(\square\:\square)&(\square\:\square)\\
\end{array}\]
where each abstraction shows with $\texttt{1}$s where in the subsequent tree of applications
its variable occurs: leaves, $\square$, are relieved of choice.
Of course, the tree under each binder determines which maps are well formed
in a highly nonlocal way: these invariants are formalised \emph{extrinsically} both in
Isabelle/HOL and in Minlog, over a context-free datatype enforcing neither scope nor shape.
Other prior art along similar lines includes the Haskell implementation by Abel and
Kraus~\cite{DBLP:journals/corr/abs-1111-0085} of a similar representation, recording at
each $\lambda$ which of the variable occurrences free below it are bound by it, in left-to-right
order, run-length encoded. Earlier still, the \emph{director strings} representation of Kennedy and Sleep,
refined by Sinot, Fernandez and Mackie~\cite{DBLP:journals/ipl/KennawayS87,DBLP:conf/rta/SinotFM03},
annotated each node with a mapping from each free variable to the set of indices of the subnodes in
which it occurs, and we shall see something similar here.

However, in this paper, we shall obtain an \emph{intrinsically} valid
representation, enforced by type, where the map information is
localized. Binding sites tell only if the variable is used; the
crucial choice points where a term comprises more than one subterm say
which variables go where, as in the director strings
representation. Not all are used in all subterms, but (as Eccles says
to Seagoon) \emph{everybody's got to be somewhere}~\cite{eccles}:
variables used nowhere have been discarded already.  This property is
delivered by a coproduct construction in the slices of the category of
order-preserving embeddings, but fear not: we shall revisit all of the
category theory required to develop the definition, especially as it
strays beyond the familiar (e.g., to Haskellers) territory of
types-and-functions.

Intrinsically well scoped de Bruijn terms date back to Bellegarde and
Hook~\cite{bellegarde.hook:substitution.monad}, using {\tt option} types to grow
a type of free variables, but hampered by lack of
polymorphic recursion in ML. Substitution (i.e., \emph{monadic} structure)
was developed for untyped terms by Bird and Paterson~\cite{bird.paterson:debruijn.nested}
and for simple types by Altenkirch and Reus~\cite{altenkirch.reus:monads.lambda}, both
dependent either on a \emph{prior} implementation of renumbering shifts (i.e., functorial
structure) or a non-structural recursion. My thesis~\cite{DBLP:phd/ethos/McBride00} follows McKinna
and Goguen~\cite{goguenmckinna} in restoring a single
structural operation abstracting `action' on variables, instantiated to
renumbering then to substitution, an approach subsequently adopted by Benton, Kennedy and
Hur~\cite{DBLP:journals/jar/BentonHKM12} and generalised to semantic actions by Allais et al.~\cite{DBLP:conf/cpp/Allais0MM17}.
Here, we go directly to substitution: \emph{shifts need no traversal}.

I present not only $\lambda$-calculus but a \emph{universe} of syntaxes inspired
by Harper, Honsell and Plotkin's
Logical Framework~\cite{DBLP:journals/jacm/HarperHP93}. I lift the \emph{sorts} of a syntax to higher
\emph{kinds}, acquiring both binding (via subterms at higher kind) and
\emph{meta}variables (at higher kind). However,
substituting a higher-kinded variable demands substitution of its parameters
\emph{hereditarily}~\cite{DBLP:conf/types/WatkinsCPW03} and \emph{simultaneously}. Thereby hangs a tale.
Abel showed how \emph{sized types} justify this process's apparently non-structural
recursion in MSFP 2006~\cite{DBLP:journals/jfp/Abel09}. As editor, I anonymised a discussion with a referee
which yielded a structural recursion for hereditary substitution of a \emph{single} variable,
instigating Keller and Altenkirch's formalization at MSFP
2010~\cite{DBLP:conf/icfp/KellerA10}. Here, at last, simultaneous hereditary substitution
becomes structurally recursive.

\section{Basic Equipment in Agda}

We shall need finite types \ensuremath{\D{Zero}}, \ensuremath{\D{One}}, and \ensuremath{\D{Two}}, named for their cardinality,
and the reflection of \ensuremath{\D{Two}} as a set of evidence for `being \ensuremath{\C{t\!t}}'.\\
\parbox{3.5in}{
\begin{hscode}\SaveRestoreHook
\column{B}{@{}>{\hspre}l<{\hspost}@{}}%
\column{9}{@{}>{\hspre}l<{\hspost}@{}}%
\column{15}{@{}>{\hspre}l<{\hspost}@{}}%
\column{E}{@{}>{\hspre}l<{\hspost}@{}}%
\>[B]{}\mathkw{data}\;{}\<[9]%
\>[9]{}\D{Zero}\;{}\<[15]%
\>[15]{}\mathbin{:}\;\D{Set}\;\mathkw{where}{}\<[E]%
\\
\>[B]{}\mathkw{record}\;{}\<[9]%
\>[9]{}\D{One}\;{}\<[15]%
\>[15]{}\mathbin{:}\;\D{Set}\;\mathkw{where}\;\mathkw{constructor}\;\C{\langle\rangle}{}\<[E]%
\\
\>[B]{}\mathkw{data}\;{}\<[9]%
\>[9]{}\D{Two}\;{}\<[15]%
\>[15]{}\mathbin{:}\;\D{Set}\;\mathkw{where}\;\C{t\!t}\;\C{f\!f}\;\mathbin{:}\;\D{Two}{}\<[E]%
\ColumnHook
\end{hscode}\resethooks
}
\vrule
\hspace*{ -0.2in}
\parbox{2.9in}{
\begin{hscode}\SaveRestoreHook
\column{B}{@{}>{\hspre}l<{\hspost}@{}}%
\column{8}{@{}>{\hspre}l<{\hspost}@{}}%
\column{E}{@{}>{\hspre}l<{\hspost}@{}}%
\>[B]{}\F{T\!t}\;\mathbin{:}\;\D{Two}\;\to \;\D{Set}{}\<[E]%
\\
\>[B]{}\F{T\!t}\;\C{t\!t}\;{}\<[8]%
\>[8]{}\mathrel{=}\;\D{One}{}\<[E]%
\\
\>[B]{}\F{T\!t}\;\C{f\!f}\;{}\<[8]%
\>[8]{}\mathrel{=}\;\D{Zero}{}\<[E]%
\ColumnHook
\end{hscode}\resethooks
}
\\
Dependent pairing is by means of the \ensuremath{\D{\Upsigma}} type, abbreviated by \ensuremath{\F{\times}} when non-dependent.
The \emph{pattern synonym} \ensuremath{\C{!}\anonymous } allows the first component to be determined by
the second: making it a right-associative prefix operator lets us write \ensuremath{\C{!}\;\C{!}\;\V{expression}}
rather than \ensuremath{\C{!}\;(\C{!}\;(\V{expression}))}.\\
\parbox{3.5in}{
\begin{hscode}\SaveRestoreHook
\column{B}{@{}>{\hspre}l<{\hspost}@{}}%
\column{3}{@{}>{\hspre}l<{\hspost}@{}}%
\column{E}{@{}>{\hspre}l<{\hspost}@{}}%
\>[B]{}\mathkw{record}\;\D{\Upsigma}\;(\V{S}\;\mathbin{:}\;\D{Set})\;(\V{T}\;\mathbin{:}\;\V{S}\;\to \;\D{Set})\;\mathbin{:}\;\D{Set}\;\mathkw{where}{}\<[E]%
\\
\>[B]{}\hsindent{3}{}\<[3]%
\>[3]{}\mathkw{constructor}\;\anonymous \C{,}\anonymous {}\<[E]%
\\
\>[B]{}\hsindent{3}{}\<[3]%
\>[3]{}\mathkw{field}\;\F{fst}\;\mathbin{:}\;\V{S};\quad\F{snd}\;\mathbin{:}\;\V{T}\;\F{fst}{}\<[E]%
\ColumnHook
\end{hscode}\resethooks
}
\vrule
\hspace*{ -0.2in}
\parbox{2.9in}{
\begin{hscode}\SaveRestoreHook
\column{B}{@{}>{\hspre}l<{\hspost}@{}}%
\column{E}{@{}>{\hspre}l<{\hspost}@{}}%
\>[B]{}\anonymous \F{\times}\anonymous \;\mathbin{:}\;\D{Set}\;\to \;\D{Set}\;\to \;\D{Set}{}\<[E]%
\\
\>[B]{}\V{S}\;\F{\times}\;\V{T}\;\mathrel{=}\;\D{\Upsigma}\;\V{S}\;\lambda \;\anonymous \;\to \;\V{T}{}\<[E]%
\\
\>[B]{}\mathkw{pattern}\;\C{!}\anonymous \;\V{t}\;\mathrel{=}\;\anonymous \;\C{,}\;\V{t}{}\<[E]%
\ColumnHook
\end{hscode}\resethooks
}

We shall also need to reason equationally. For all its imperfections in matters of
\emph{extensionality}, it will be convenient to define equality inductively, enabling the \ensuremath{\mathkw{rewrite}} construct
in equational proofs.
\begin{hscode}\SaveRestoreHook
\column{B}{@{}>{\hspre}l<{\hspost}@{}}%
\column{E}{@{}>{\hspre}l<{\hspost}@{}}%
\>[B]{}\mathkw{data}\;\anonymous \D{=\!\!\!\!=}\anonymous \;\{\mskip1.5mu \V{X}\;\mathbin{:}\;\D{Set}\mskip1.5mu\}\;(\V{x}\;\mathbin{:}\;\V{X})\;\mathbin{:}\;\V{X}\;\to \;\D{Set}\;\mathkw{where}\;\C{refl}\;\mathbin{:}\;\V{x}\;\D{=\!\!\!\!=}\;\V{x}{}\<[E]%
\ColumnHook
\end{hscode}\resethooks

\section{$\OPE{\ensuremath{\V{K}}}$: The (Semi-Simplicial) Category of Order-Preserving Embeddings}

No category theorist would mistake me for one of their own. However, the key technology
in this paper can be helpfully conceptualised categorically. Category theory is just the
study of compositionality --- for everything, not just sets-and-functions. Here, we have an
opportunity to develop categorical structure away from the usual apparatus for
programming with functions. Let us therefore revisit the basics.

\paragraph{Category (I): Objects and Morphisms.~} A \emph{category} is given by a class of \emph{objects} and a family
of \emph{morphisms} (or \emph{arrows}) indexed by two objects: \emph{source} and
\emph{target}. Abstractly, we may write $\mathbb{C}$ for a given
category, $|\mathbb{C}|$ for its objects, and
$\mathbb{C}(S,T)$ for its morphisms with given source and target,
$S,T\in|\mathbb{C}|$.

The rest will follow, but let
us fix these notions for our example category, $\OPE{\ensuremath{\V{K}}}$, of
\emph{order-preserving embeddings} between variable \emph{scopes}.
Objects are given as backward (or `snoc') lists of the \emph{kinds}, \ensuremath{\V{K}}, of
variables. (I habitually suppress \ensuremath{\V{K}} and just write
$\OPE{}$ for the category.)
Backward lists respect the tradition of writing contexts
left of judgements in rules and growing them rightwards.
However, I say `scope' rather than `context':
we track variable availability, but perhaps
not all contextual data. Moreover, I take the typesetting liberty
of hiding inferable prefixes of implicit quantifiers,
to reduce clutter.\\
\newcommand{\apo}{^\prime}
\begin{tabular}{@{}l|l@{}}
\raisebox{0.9in}[0.9in][0in]{\parbox[t]{2.7in}{
\begin{hscode}\SaveRestoreHook
\column{B}{@{}>{\hspre}l<{\hspost}@{}}%
\column{3}{@{}>{\hspre}l<{\hspost}@{}}%
\column{8}{@{}>{\hspre}l<{\hspost}@{}}%
\column{E}{@{}>{\hspre}l<{\hspost}@{}}%
\>[B]{}\mathkw{data}\;\D{Bwd}\;(\V{K}\;\mathbin{:}\;\D{Set})\;\mathbin{:}\;\D{Set}\;\mathkw{where}{}\<[E]%
\\
\>[B]{}\hsindent{3}{}\<[3]%
\>[3]{}\anonymous \C{\mbox{{}-}\!,}\anonymous \;{}\<[8]%
\>[8]{}\mathbin{:}\;\D{Bwd}\;\V{K}\;\to \;\V{K}\;\to \;\D{Bwd}\;\V{K}{}\<[E]%
\\
\>[B]{}\hsindent{3}{}\<[3]%
\>[3]{}\C{[]}\;{}\<[8]%
\>[8]{}\mathbin{:}\;\D{Bwd}\;\V{K}{}\<[E]%
\ColumnHook
\end{hscode}\resethooks
}}
&
\hspace*{ -0.3in}
\raisebox{0.9in}[0.9in][0in]{\parbox[t]{3in}{
\begin{hscode}\SaveRestoreHook
\column{B}{@{}>{\hspre}l<{\hspost}@{}}%
\column{3}{@{}>{\hspre}l<{\hspost}@{}}%
\column{8}{@{}>{\hspre}l<{\hspost}@{}}%
\column{25}{@{}>{\hspre}l<{\hspost}@{}}%
\column{41}{@{}>{\hspre}l<{\hspost}@{}}%
\column{54}{@{}>{\hspre}c<{\hspost}@{}}%
\column{54E}{@{}l@{}}%
\column{57}{@{}>{\hspre}l<{\hspost}@{}}%
\column{61}{@{}>{\hspre}l<{\hspost}@{}}%
\column{65}{@{}>{\hspre}l<{\hspost}@{}}%
\column{69}{@{}>{\hspre}l<{\hspost}@{}}%
\column{73}{@{}>{\hspre}l<{\hspost}@{}}%
\column{78}{@{}>{\hspre}l<{\hspost}@{}}%
\column{82}{@{}>{\hspre}l<{\hspost}@{}}%
\column{E}{@{}>{\hspre}l<{\hspost}@{}}%
\>[B]{}\mathkw{data}\;\anonymous \D{\sqsubseteq}\anonymous \;{}\<[25]%
\>[25]{}\mathbin{:}\;\D{Bwd}\;\V{K}\;\to \;\D{Bwd}\;\V{K}\;\to \;\D{Set}\;\mathkw{where}{}\<[E]%
\\
\>[B]{}\hsindent{3}{}\<[3]%
\>[3]{}\anonymous \C{o\apo}\;{}\<[8]%
\>[8]{}\mathbin{:}\;{}\<[41]%
\>[41]{}\V{iz}\;\D{\sqsubseteq}\;\V{jz}\;\to \;{}\<[57]%
\>[57]{}\V{iz}\;{}\<[69]%
\>[69]{}\D{\sqsubseteq}\;{}\<[73]%
\>[73]{}(\V{jz}\;{}\<[78]%
\>[78]{}\C{\mbox{{}-}\!,}\;{}\<[82]%
\>[82]{}\V{k}){}\<[E]%
\\
\>[B]{}\hsindent{3}{}\<[3]%
\>[3]{}\anonymous \C{os}\;{}\<[8]%
\>[8]{}\mathbin{:}\;{}\<[41]%
\>[41]{}\V{iz}\;\D{\sqsubseteq}\;\V{jz}\;\to \;{}\<[54]%
\>[54]{}({}\<[54E]%
\>[57]{}\V{iz}\;{}\<[61]%
\>[61]{}\C{\mbox{{}-}\!,}\;{}\<[65]%
\>[65]{}\V{k})\;{}\<[69]%
\>[69]{}\D{\sqsubseteq}\;{}\<[73]%
\>[73]{}(\V{jz}\;{}\<[78]%
\>[78]{}\C{\mbox{{}-}\!,}\;{}\<[82]%
\>[82]{}\V{k}){}\<[E]%
\\
\>[B]{}\hsindent{3}{}\<[3]%
\>[3]{}\C{oz}\;{}\<[8]%
\>[8]{}\mathbin{:}\;{}\<[61]%
\>[61]{}\C{[]}\;{}\<[69]%
\>[69]{}\D{\sqsubseteq}\;{}\<[78]%
\>[78]{}\C{[]}{}\<[E]%
\ColumnHook
\end{hscode}\resethooks
}}
\end{tabular}

The morphisms, \ensuremath{\V{iz}\;\D{\sqsubseteq}\;\V{jz}}, of $\OPE{}$ embed a source
into a target scope. Colloquially, we may call them
`thinnings', as they dilute the variables of the source scope with
more. I write step constructors postfix,
so thinnings (like scopes) grow on the right. 
Now, where I give myself away as a type theorist is that I do not
consider the notion of `morphism' to make sense without prior source and
target objects. The type \ensuremath{\V{iz}\;\D{\sqsubseteq}\;\V{jz}} (which is a little more mnemonic than
$\OPE{}(\ensuremath{\V{iz}},\ensuremath{\V{jz}})$) is the type of `thinnings from \ensuremath{\V{iz}} to
\ensuremath{\V{jz}}': there is no type of `thinnings' \emph{per se}.

Altenkirch, Hofmann and Streicher~\cite{DBLP:conf/ctcs/AltenkirchHS95},
from whom I learned this notion, take the dual view of morphisms as
\emph{selecting} one subcontext from another. When \ensuremath{\V{K}\;\mathrel{=}\;\D{One}},
objects represent numbers and \ensuremath{\D{\sqsubseteq}} generates Pascal's
Triangle; excluding the empty scope and allowing \emph{degenerate}
(non-injective) maps yields $\Delta$, the \emph{simplex} category beloved of topologists.

Let us have an example thinning: here, we embed a scope with three variables
into a scope with five.
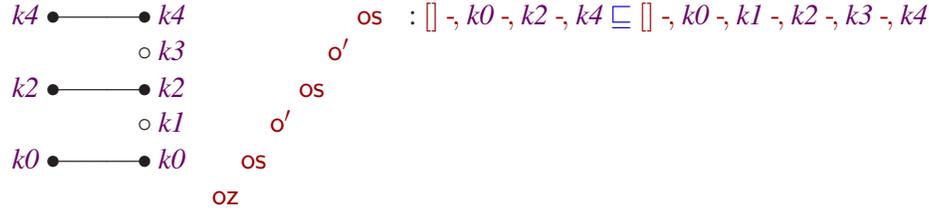
\[\begin{array}{r@{\!\!}c@{\!\!}lll}
\ensuremath{\V{k4}}\;\bullet & -\!\!\!-\!\!\!-\!\!\!-\!\!\!- & \bullet\;\ensuremath{\V{k4}} & \quad\quad\quad\quad\quad\ensuremath{\C{os}} & \ensuremath{\mathbin{:}\;\C{[]}\;\C{\mbox{{}-}\!,}\;\V{k0}\;\C{\mbox{{}-}\!,}\;\V{k2}\;\C{\mbox{{}-}\!,}\;\V{k4}\;\D{\sqsubseteq}\;\C{[]}\;\C{\mbox{{}-}\!,}\;\V{k0}\;\C{\mbox{{}-}\!,}\;\V{k1}\;\C{\mbox{{}-}\!,}\;\V{k2}\;\C{\mbox{{}-}\!,}\;\V{k3}\;\C{\mbox{{}-}\!,}\;\V{k4}} \\
             &                  & \circ\;\ensuremath{\V{k3}} & \quad\quad\quad\quad\ensuremath{\C{o\apo}}\\
\ensuremath{\V{k2}}\;\bullet & -\!\!\!-\!\!\!-\!\!\!-\!\!\!- & \bullet\;\ensuremath{\V{k2}} & \quad\quad\quad\ensuremath{\C{os}} \\
             &                  & \circ\;\ensuremath{\V{k1}} & \quad\quad\ensuremath{\C{o\apo}}\\
\ensuremath{\V{k0}}\;\bullet & -\!\!\!-\!\!\!-\!\!\!-\!\!\!- & \bullet\;\ensuremath{\V{k0}}& \quad\ensuremath{\C{os}}\\
&&& \ensuremath{\C{oz}}
\end{array}\]

\paragraph{Category (II): Identity and Composition.~} In any category, certain morphisms
must exist. Each object $X\in|\mathbb{C}|$
has an \emph{identity} $\iota_X\in\mathbb{C}(X,X)$, and wherever the
target of one morphism meets the source of another, their \emph{composite}
makes a direct path:
if $f\in\mathbb{C}(R,S)$ and $g\in\mathbb{C}(S,T)$, then $(f;g)\in\mathbb{C}(R,T)$.

E.g., every scope has the identity thinning, \ensuremath{\F{oi}}, and thinnings compose via \ensuremath{\F{\fatsemi}}.
(For functions, it is usual to write $g\cdot f$ for `$g$ \emph{after} $f$' rather than $f;g$ for
`$f$ \emph{then} $g$', but for thinnings I retain spatial
intuition.)
\\
\begin{tabular}{@{}l|l@{}}
\raisebox{1.1in}[1in][0in]{\parbox[t]{3.1in}{
\begin{hscode}\SaveRestoreHook
\column{B}{@{}>{\hspre}l<{\hspost}@{}}%
\column{19}{@{}>{\hspre}l<{\hspost}@{}}%
\column{45}{@{}>{\hspre}l<{\hspost}@{}}%
\column{E}{@{}>{\hspre}l<{\hspost}@{}}%
\>[B]{}\F{oi}\;\mathbin{:}\;{}\<[45]%
\>[45]{}\V{kz}\;\D{\sqsubseteq}\;\V{kz}{}\<[E]%
\\
\>[B]{}\F{oi}\;\{\mskip1.5mu \V{kz}\;\mathrel{=}\;\V{iz}\;\C{\mbox{{}-}\!,}\;\V{k}\mskip1.5mu\}\;{}\<[19]%
\>[19]{}\mathrel{=}\;\F{oi}\;\C{os}\mbox{\onelinecomment  \ensuremath{\C{os}} preserves \ensuremath{\F{oi}}}{}\<[E]%
\\
\>[B]{}\F{oi}\;\{\mskip1.5mu \V{kz}\;\mathrel{=}\;\C{[]}\mskip1.5mu\}\;{}\<[19]%
\>[19]{}\mathrel{=}\;\C{oz}{}\<[E]%
\ColumnHook
\end{hscode}\resethooks
}}
& \hspace*{ -0.3in}
\raisebox{1.1in}[1in][0in]{\parbox[t]{3in}{
\begin{hscode}\SaveRestoreHook
\column{B}{@{}>{\hspre}l<{\hspost}@{}}%
\column{8}{@{}>{\hspre}l<{\hspost}@{}}%
\column{19}{@{}>{\hspre}l<{\hspost}@{}}%
\column{55}{@{}>{\hspre}l<{\hspost}@{}}%
\column{E}{@{}>{\hspre}l<{\hspost}@{}}%
\>[B]{}\anonymous \F{\fatsemi}\anonymous \;\mathbin{:}\;{}\<[55]%
\>[55]{}\V{iz}\;\D{\sqsubseteq}\;\V{jz}\;\to \;\V{jz}\;\D{\sqsubseteq}\;\V{kz}\;\to \;\V{iz}\;\D{\sqsubseteq}\;\V{kz}{}\<[E]%
\\
\>[B]{}\V{\theta}\;{}\<[8]%
\>[8]{}\F{\fatsemi}\;\V{\phi}\;\C{o\apo}\;{}\<[19]%
\>[19]{}\mathrel{=}\;(\V{\theta}\;\F{\fatsemi}\;\V{\phi})\;\C{o\apo}{}\<[E]%
\\
\>[B]{}\V{\theta}\;\C{o\apo}\;{}\<[8]%
\>[8]{}\F{\fatsemi}\;\V{\phi}\;\C{os}\;{}\<[19]%
\>[19]{}\mathrel{=}\;(\V{\theta}\;\F{\fatsemi}\;\V{\phi})\;\C{o\apo}{}\<[E]%
\\
\>[B]{}\V{\theta}\;\C{os}\;{}\<[8]%
\>[8]{}\F{\fatsemi}\;\V{\phi}\;\C{os}\;{}\<[19]%
\>[19]{}\mathrel{=}\;(\V{\theta}\;\F{\fatsemi}\;\V{\phi})\;\C{os}\mbox{\onelinecomment  \ensuremath{\C{os}} preserves \ensuremath{\F{\fatsemi}}}{}\<[E]%
\\
\>[B]{}\C{oz}\;{}\<[8]%
\>[8]{}\F{\fatsemi}\;\C{oz}\;{}\<[19]%
\>[19]{}\mathrel{=}\;\C{oz}{}\<[E]%
\ColumnHook
\end{hscode}\resethooks
}}
\end{tabular}\vspace*{ 0.05in}

By way of example, let us plot specific uses of identity and composition.
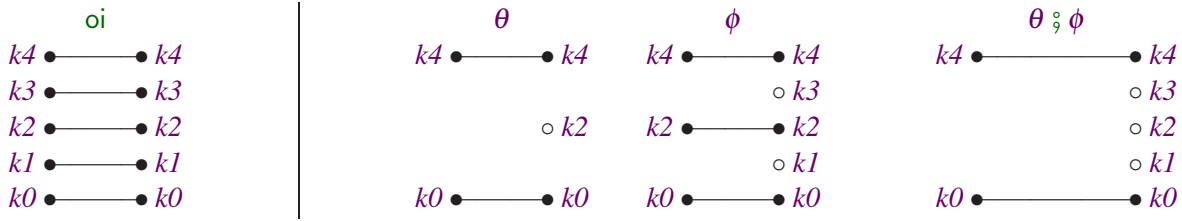
\[\begin{array}{r@{\!\!}c@{\!\!}l@{\qquad\qquad}|@{\qquad\qquad}r@{\!\!}c@{\!\!}l@{\qquad}r@{\!\!}c@{\!\!}l@{\qquad\qquad}r@{\!\!}c@{\!\!}l}
&\ensuremath{\F{oi}}&& &\ensuremath{\V{\theta}}&& &\ensuremath{\V{\phi}}&& &\ensuremath{\V{\theta}\;\F{\fatsemi}\;\V{\phi}}& \\
\ensuremath{\V{k4}}\;\bullet & -\!\!\!-\!\!\!-\!\!\!-\!\!\!- & \bullet\;\ensuremath{\V{k4}} &
\ensuremath{\V{k4}}\;\bullet & -\!\!\!-\!\!\!-\!\!\!-\!\!\!- & \bullet\;\ensuremath{\V{k4}} &
\ensuremath{\V{k4}}\;\bullet & -\!\!\!-\!\!\!-\!\!\!-\!\!\!- & \bullet\;\ensuremath{\V{k4}} &
\ensuremath{\V{k4}}\;\bullet & -\!\!\!-\!\!\!-\!\!\!-\!\!\!-\!\!\!-\!\!\!-\!\!\!-\!\!\!- & \bullet\;\ensuremath{\V{k4}} \\
\ensuremath{\V{k3}}\;\bullet & -\!\!\!-\!\!\!-\!\!\!-\!\!\!- & \bullet\;\ensuremath{\V{k3}} &
&&&          &                  & \circ\;\ensuremath{\V{k3}}&         &                  & \circ\;\ensuremath{\V{k3}} \\
\ensuremath{\V{k2}}\;\bullet & -\!\!\!-\!\!\!-\!\!\!-\!\!\!- & \bullet\;\ensuremath{\V{k2}} &
&& \circ\;\ensuremath{\V{k2}} &          
\ensuremath{\V{k2}}\;\bullet & -\!\!\!-\!\!\!-\!\!\!-\!\!\!- & \bullet\;\ensuremath{\V{k2}} &&& \circ\;\ensuremath{\V{k2}} \\
\ensuremath{\V{k1}}\;\bullet & -\!\!\!-\!\!\!-\!\!\!-\!\!\!- & \bullet\;\ensuremath{\V{k1}} &
&&&          &                  & \circ\;\ensuremath{\V{k1}}&          &                  & \circ\;\ensuremath{\V{k1}} \\
\ensuremath{\V{k0}}\;\bullet & -\!\!\!-\!\!\!-\!\!\!-\!\!\!- & \bullet\;\ensuremath{\V{k0}} &
\ensuremath{\V{k0}}\;\bullet & -\!\!\!-\!\!\!-\!\!\!-\!\!\!- & \bullet\;\ensuremath{\V{k0}}&
\ensuremath{\V{k0}}\;\bullet & -\!\!\!-\!\!\!-\!\!\!-\!\!\!- & \bullet\;\ensuremath{\V{k0}}&
\ensuremath{\V{k0}}\;\bullet & -\!\!\!-\!\!\!-\!\!\!-\!\!\!-\!\!\!-\!\!\!-\!\!\!-\!\!\!- & \bullet\;\ensuremath{\V{k0}}\\
\end{array}\]

\paragraph{Category (III): Laws.~} To complete the definition of a
category, we must say which laws are satisfied by identity and
composition. Composition \emph{absorbs} identity on the left and on
the right. Moreover, composition is \emph{associative}, meaning that
any sequence of morphisms which fit together target-to-source can be
composed without the specific pairwise grouping choices making a
difference. That is, we have three laws which are presented as
\emph{equations}, at which point any type theorist will want to know
what is meant by `equal': I shall always be careful to say.
Our thinnings are first-order, so \ensuremath{\D{=\!\!\!\!=}} will serve.
With this definition in place, we may then state the laws. I
omit the proofs, which go by functional induction.
\[
\ensuremath{\F{law-}\F{oi}\F{\fatsemi}\;\mathbin{:}\;\F{oi}\;\F{\fatsemi}\;\V{\theta}\;\D{=\!\!\!\!=}\;\V{\theta}} \qquad
\ensuremath{\F{law-}\F{\fatsemi}\F{oi}\;\mathbin{:}\;\V{\theta}\;\F{\fatsemi}\;\F{oi}\;\D{=\!\!\!\!=}\;\V{\theta}} \qquad
\ensuremath{\F{law-}\F{\fatsemi}\F{\fatsemi}\;\mathbin{:}\;\V{\theta}\;\F{\fatsemi}\;(\V{\phi}\;\F{\fatsemi}\;\V{\psi})\;\D{=\!\!\!\!=}\;(\V{\theta}\;\F{\fatsemi}\;\V{\phi})\;\F{\fatsemi}\;\V{\psi}}
\]

As one might expect, order-preserving embeddings have a strong
antisymmetry property that one cannot expect of categories in general.
The \emph{only} invertible arrows are the identities.
Note that we must match on the proof of \ensuremath{\V{iz}\;\D{=\!\!\!\!=}\;\V{jz}} even to claim
that \ensuremath{\V{\theta}} and \ensuremath{\V{\phi}} are the identity.
\begin{hscode}\SaveRestoreHook
\column{B}{@{}>{\hspre}l<{\hspost}@{}}%
\column{49}{@{}>{\hspre}l<{\hspost}@{}}%
\column{E}{@{}>{\hspre}l<{\hspost}@{}}%
\>[B]{}\F{antisym}\;\mathbin{:}\;{}\<[49]%
\>[49]{}(\V{\theta}\;\mathbin{:}\;\V{iz}\;\D{\sqsubseteq}\;\V{jz})\;(\V{\phi}\;\mathbin{:}\;\V{jz}\;\D{\sqsubseteq}\;\V{iz})\;\to \;\D{\Upsigma}\;(\V{iz}\;\D{=\!\!\!\!=}\;\V{jz})\;\lambda \;\{\mskip1.5mu \C{refl}\;\to \;(\V{\theta}\;\D{=\!\!\!\!=}\;\F{oi})\;\F{\times}\;(\V{\phi}\;\D{=\!\!\!\!=}\;\F{oi})\mskip1.5mu\}{}\<[E]%
\ColumnHook
\end{hscode}\resethooks

\paragraph{Example: de Bruijn Syntax via $\OPE{\ensuremath{\D{One}}}$.}~
De Bruijn indices are numbers~\cite{deBruijn:dummies}, perhaps
a type-enforced bound~\cite{bellegarde.hook:substitution.monad,bird.paterson:debruijn.nested,altenkirch.reus:monads.lambda}.
Singleton thinnings, \ensuremath{\V{k}\;\F{\leftarrow}\;\V{kz}\;\mathrel{=}\;\C{[]}\;\C{\mbox{{}-}\!,}\;\V{k}\;\D{\sqsubseteq}\;\V{kz}}, can play this r\^ole in a syntax.
 \\
\begin{tabular}{@{}l|l@{}}
\raisebox{0.9in}[0.8in][0in]{\parbox[t]{3in}{
\begin{hscode}\SaveRestoreHook
\column{B}{@{}>{\hspre}l<{\hspost}@{}}%
\column{3}{@{}>{\hspre}l<{\hspost}@{}}%
\column{8}{@{}>{\hspre}l<{\hspost}@{}}%
\column{26}{@{}>{\hspre}l<{\hspost}@{}}%
\column{37}{@{}>{\hspre}l<{\hspost}@{}}%
\column{E}{@{}>{\hspre}l<{\hspost}@{}}%
\>[B]{}\mathkw{data}\;\D{Lam}\;(\V{iz}\;\mathbin{:}\;\D{Bwd}\;\D{One})\;\mathbin{:}\;\D{Set}\;\mathkw{where}{}\<[E]%
\\
\>[B]{}\hsindent{3}{}\<[3]%
\>[3]{}\C{\scriptstyle\#}\;{}\<[8]%
\>[8]{}\mathbin{:}\;(\V{x}\;\mathbin{:}\;\C{\langle\rangle}\;\F{\leftarrow}\;\V{iz})\;{}\<[26]%
\>[26]{}\to \;{}\<[37]%
\>[37]{}\D{Lam}\;\V{iz}{}\<[E]%
\\
\>[B]{}\hsindent{3}{}\<[3]%
\>[3]{}\anonymous \C{\scriptstyle\$}\anonymous \;{}\<[8]%
\>[8]{}\mathbin{:}\;(\V{f}\;\V{s}\;\mathbin{:}\;\D{Lam}\;\V{iz})\;\to \;{}\<[37]%
\>[37]{}\D{Lam}\;\V{iz}{}\<[E]%
\\
\>[B]{}\hsindent{3}{}\<[3]%
\>[3]{}\C{\uplambda}\;{}\<[8]%
\>[8]{}\mathbin{:}\;(\V{t}\;\mathbin{:}\;\D{Lam}\;(\V{iz}\;\C{\mbox{{}-}\!,}\;\C{\langle\rangle}))\;\to \;{}\<[37]%
\>[37]{}\D{Lam}\;\V{iz}{}\<[E]%
\ColumnHook
\end{hscode}\resethooks
}}
&
\hspace*{ -0.3in}
\raisebox{0.9in}[0.8in][0in]{\parbox[t]{2.5in}{
\begin{hscode}\SaveRestoreHook
\column{B}{@{}>{\hspre}l<{\hspost}@{}}%
\column{10}{@{}>{\hspre}l<{\hspost}@{}}%
\column{38}{@{}>{\hspre}l<{\hspost}@{}}%
\column{E}{@{}>{\hspre}l<{\hspost}@{}}%
\>[B]{}\anonymous \F{\uparrow}\anonymous \;\mathbin{:}\;{}\<[38]%
\>[38]{}\D{Lam}\;\V{iz}\;\to \;\V{iz}\;\D{\sqsubseteq}\;\V{jz}\;\to \;\D{Lam}\;\V{jz}{}\<[E]%
\\
\>[B]{}\C{\scriptstyle\#}\;\V{i}\;{}\<[10]%
\>[10]{}\F{\uparrow}\;\V{\theta}\;\mathrel{=}\;\C{\scriptstyle\#}\;(\V{i}\;\F{\fatsemi}\;\V{\theta}){}\<[E]%
\\
\>[B]{}(\V{f}\;\C{\scriptstyle\$}\;\V{s})\;{}\<[10]%
\>[10]{}\F{\uparrow}\;\V{\theta}\;\mathrel{=}\;(\V{f}\;\F{\uparrow}\;\V{\theta})\;\C{\scriptstyle\$}\;(\V{s}\;\F{\uparrow}\;\V{\theta}){}\<[E]%
\\
\>[B]{}\C{\uplambda}\;\V{t}\;{}\<[10]%
\>[10]{}\F{\uparrow}\;\V{\theta}\;\mathrel{=}\;\C{\uplambda}\;(\V{t}\;\F{\uparrow}\;\V{\theta}\;\C{os}){}\<[E]%
\ColumnHook
\end{hscode}\resethooks
}}
\end{tabular}

Variables are represented by pointing, eliminating redundant choice of names,
but it is only when we point to one variable that we exclude
the others. Thus de Bruijn indexing
effectively uses thinnings to discard unwanted variables as
\emph{late} as possible, in the \emph{leaves} of syntax trees.

Note how the scope index \ensuremath{\V{iz}} is the
target of a thinning in \ensuremath{\C{\scriptstyle\#}} and weakened in \ensuremath{\C{\uplambda}}.
Hence, thinnings act on terms ultimately
by postcomposition, but because terms keep their thinnings at their leaves,
we must hunt the entire tree to find them.
Now consider the other canonical placement of
thinnings, nearest the \emph{root}, discarding unused variables
as \emph{early} as possible.

\section{Slices of Thinnings}

If we fix the target of thinnings, \ensuremath{(\anonymous \D{\sqsubseteq}\;\V{kz})}, we obtain the notion of
\emph{subscopes} of a given \ensuremath{\V{kz}}. Fixing a target is a standard way
to construct a new category whose objects are given by
morphisms of the original.
\vspace*{ -0.2in} \\
\parbox[t]{4in}{
\paragraph{Slice Category.~} If
$\mathbb{C}$ is a category and $I$ one of its objects, the \emph{slice category}
$\mathbb{C}/I$ has as its objects pairs $(S, f)$, where $S$ is an object
of $\mathbb{C}$ and $f : S \to I$ is a morphism in $\mathbb{C}$. A morphism
in $(\mathbb{C}/I)((S, f),(T, g))$ is some $h : S \to T$ such that $f = h;g$.
(The dotted regions in the diagram show the objects in the slice.)
}
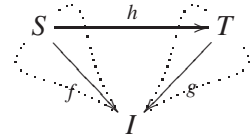
\parbox[t]{2in}{
\[\xymatrix{
  S \ar[rr]^h \ar[dr]_f &   & T \ar[dl]^g \\
    & I \ar@{.}@(ul,dl)[ul]\ar@{.}@(ul,ur)[ul]
        \ar@{.}@(ur,dr)[ur]\ar@{.}@(ur,ul)[ur]
        &
}\]}

~

That is, the morphisms are \emph{triangles}. A seasoned dependently typed
programmer will be nervous at a definition like the following
(where the \ensuremath{\anonymous } after \ensuremath{\D{\Upsigma}} asks Agda to compute the type \ensuremath{\V{iz}\;\D{\sqsubseteq}\;\V{jz}} of \ensuremath{\V{\theta}}):
\begin{hscode}\SaveRestoreHook
\column{B}{@{}>{\hspre}l<{\hspost}@{}}%
\column{45}{@{}>{\hspre}l<{\hspost}@{}}%
\column{E}{@{}>{\hspre}l<{\hspost}@{}}%
\>[B]{}\V{\psi}\;\F{\to_{\slash}}\;\V{\phi}\;\mathrel{=}\;\D{\Upsigma}\;\anonymous \;\lambda \;\V{\theta}\;\to \;(\V{\theta}\;\F{\fatsemi}\;\V{\phi})\;\D{=\!\!\!\!=}\;\V{\psi}{}\<[45]%
\>[45]{}\mbox{\onelinecomment  beware of \ensuremath{\F{\fatsemi}}!}{}\<[E]%
\ColumnHook
\end{hscode}\resethooks
because the equation restricts us when it comes to manipulating
triangles. Dependent pattern matching relies on \emph{unification} of
indices, but defined functions like \ensuremath{\F{\fatsemi}} make unification
difficult, obliging us to reason about the \emph{edges} of the triangles.
It helps at this point to define the \emph{graph} of \ensuremath{\F{\fatsemi}} inductively.
\\
\begin{tabular}{@{}l|l@{}}
\raisebox{1.1in}[1in][0in]{\parbox[t]{3.8in}{
\begin{hscode}\SaveRestoreHook
\column{B}{@{}>{\hspre}l<{\hspost}@{}}%
\column{3}{@{}>{\hspre}l<{\hspost}@{}}%
\column{10}{@{}>{\hspre}l<{\hspost}@{}}%
\column{24}{@{}>{\hspre}l<{\hspost}@{}}%
\column{58}{@{}>{\hspre}l<{\hspost}@{}}%
\column{91}{@{}>{\hspre}l<{\hspost}@{}}%
\column{107}{@{}>{\hspre}l<{\hspost}@{}}%
\column{134}{@{}>{\hspre}c<{\hspost}@{}}%
\column{134E}{@{}l@{}}%
\column{137}{@{}>{\hspre}l<{\hspost}@{}}%
\column{141}{@{}>{\hspre}l<{\hspost}@{}}%
\column{146}{@{}>{\hspre}l<{\hspost}@{}}%
\column{151}{@{}>{\hspre}l<{\hspost}@{}}%
\column{156}{@{}>{\hspre}l<{\hspost}@{}}%
\column{161}{@{}>{\hspre}l<{\hspost}@{}}%
\column{E}{@{}>{\hspre}l<{\hspost}@{}}%
\>[B]{}\mathkw{data}\;\D{Tri}\;{}\<[24]%
\>[24]{}\mathbin{:}\;{}\<[58]%
\>[58]{}\V{iz}\;\D{\sqsubseteq}\;\V{jz}\;\to \;\V{jz}\;\D{\sqsubseteq}\;\V{kz}\;\to \;\V{iz}\;\D{\sqsubseteq}\;\V{kz}\;\to \;\D{Set}\;\mathkw{where}{}\<[E]%
\\
\>[B]{}\hsindent{3}{}\<[3]%
\>[3]{}\anonymous \C{t\mbox{{}-}\apo\!\apo}\;{}\<[10]%
\>[10]{}\mathbin{:}\;{}\<[91]%
\>[91]{}\D{Tri}\;\V{\theta}\;\V{\phi}\;\V{\psi}\;\to \;\D{Tri}\;{}\<[137]%
\>[137]{}\V{\theta}\;{}\<[146]%
\>[146]{}(\V{\phi}\;{}\<[151]%
\>[151]{}\C{o\apo})\;{}\<[156]%
\>[156]{}(\V{\psi}\;{}\<[161]%
\>[161]{}\C{o\apo}){}\<[E]%
\\
\>[B]{}\hsindent{3}{}\<[3]%
\>[3]{}\anonymous \C{t\apo\!s\apo}\;{}\<[10]%
\>[10]{}\mathbin{:}\;{}\<[91]%
\>[91]{}\D{Tri}\;\V{\theta}\;\V{\phi}\;\V{\psi}\;\to \;\D{Tri}\;{}\<[134]%
\>[134]{}({}\<[134E]%
\>[137]{}\V{\theta}\;{}\<[141]%
\>[141]{}\C{o\apo})\;{}\<[146]%
\>[146]{}(\V{\phi}\;{}\<[151]%
\>[151]{}\C{os})\;{}\<[156]%
\>[156]{}(\V{\psi}\;{}\<[161]%
\>[161]{}\C{o\apo}){}\<[E]%
\\
\>[B]{}\hsindent{3}{}\<[3]%
\>[3]{}\anonymous \C{tsss}\;{}\<[10]%
\>[10]{}\mathbin{:}\;{}\<[91]%
\>[91]{}\D{Tri}\;\V{\theta}\;\V{\phi}\;\V{\psi}\;\to \;\D{Tri}\;{}\<[134]%
\>[134]{}({}\<[134E]%
\>[137]{}\V{\theta}\;{}\<[141]%
\>[141]{}\C{os})\;{}\<[146]%
\>[146]{}(\V{\phi}\;{}\<[151]%
\>[151]{}\C{os})\;{}\<[156]%
\>[156]{}(\V{\psi}\;{}\<[161]%
\>[161]{}\C{os}){}\<[E]%
\\
\>[B]{}\hsindent{3}{}\<[3]%
\>[3]{}\C{tzzz}\;{}\<[10]%
\>[10]{}\mathbin{:}\;{}\<[107]%
\>[107]{}\D{Tri}\;{}\<[141]%
\>[141]{}\C{oz}\;{}\<[151]%
\>[151]{}\C{oz}\;{}\<[161]%
\>[161]{}\C{oz}{}\<[E]%
\ColumnHook
\end{hscode}\resethooks
}}
& \hspace*{ -0.4in}
\raisebox{1.1in}[1in][0in]{\parbox[t]{2.2in}{
\begin{hscode}\SaveRestoreHook
\column{B}{@{}>{\hspre}l<{\hspost}@{}}%
\column{7}{@{}>{\hspre}l<{\hspost}@{}}%
\column{98}{@{}>{\hspre}l<{\hspost}@{}}%
\column{E}{@{}>{\hspre}l<{\hspost}@{}}%
\>[B]{}\F{tri}\;{}\<[7]%
\>[7]{}\mathbin{:}\;{}\<[98]%
\>[98]{}(\V{\theta}\;\mathbin{:}\;\V{iz}\;\D{\sqsubseteq}\;\V{jz})\;(\V{\phi}\;\mathbin{:}\;\V{jz}\;\D{\sqsubseteq}\;\V{kz})\;\to \;{}\<[E]%
\\
\>[98]{}\D{Tri}\;\V{\theta}\;\V{\phi}\;(\V{\theta}\;\F{\fatsemi}\;\V{\phi}){}\<[E]%
\\[\blanklineskip]%
\>[B]{}\F{comp}\;{}\<[7]%
\>[7]{}\mathbin{:}\;{}\<[98]%
\>[98]{}\D{Tri}\;\V{\theta}\;\V{\phi}\;\V{\psi}\;\to \;\V{\psi}\;\D{=\!\!\!\!=}\;(\V{\theta}\;\F{\fatsemi}\;\V{\phi}){}\<[E]%
\ColumnHook
\end{hscode}\resethooks
}}
\end{tabular}

\noindent
The indexing is entirely in constructor form, which will allow
easy unification. Moreover, all the \emph{data} in a \ensuremath{\D{Tri}} structure
come from its \emph{indices}.
Easy inductions show that \ensuremath{\D{Tri}} is precisely the graph of \ensuremath{\F{\fatsemi}}.

The example composition given above can be rendered a triangle, as follows:
\begin{hscode}\SaveRestoreHook
\column{B}{@{}>{\hspre}l<{\hspost}@{}}%
\column{55}{@{}>{\hspre}l<{\hspost}@{}}%
\column{E}{@{}>{\hspre}l<{\hspost}@{}}%
\>[B]{}\F{egTri}\;\mathbin{:}\;{}\<[55]%
\>[55]{}\D{Tri}\;\{\mskip1.5mu \V{kz}\;\mathrel{=}\;\C{[]}\;\C{\mbox{{}-}\!,}\;\V{k0}\;\C{\mbox{{}-}\!,}\;\V{k1}\;\C{\mbox{{}-}\!,}\;\V{k2}\;\C{\mbox{{}-}\!,}\;\V{k3}\;\C{\mbox{{}-}\!,}\;\V{k4}\mskip1.5mu\}\;(\C{oz}\;\C{os}\;\C{o\apo}\;\C{os})\;(\C{oz}\;\C{os}\;\C{o\apo}\;\C{os}\;\C{o\apo}\;\C{os})\;(\C{oz}\;\C{os}\;\C{o\apo}\;\C{o\apo}\;\C{o\apo}\;\C{os}){}\<[E]%
\\
\>[B]{}\F{egTri}\;\mathrel{=}\;\C{tzzz}\;\C{tsss}\;\C{t\mbox{{}-}\apo\!\apo}\;\C{t\apo\!s\apo}\;\C{t\mbox{{}-}\apo\!\apo}\;\C{tsss}{}\<[E]%
\ColumnHook
\end{hscode}\resethooks

Morphisms in the slice can now be triangles:~ \ensuremath{\V{\psi}\;\F{\to_{\slash}}\;\V{\phi}\;\mathrel{=}\;\D{\Upsigma}\;\anonymous \;\lambda \;\V{\theta}\;\to \;\D{Tri}\;\V{\theta}\;\V{\phi}\;\V{\psi}}.

A useful $\OPE{}$-specific property is that morphisms in $\OPE{}/\ensuremath{\V{kz}}$ are \emph{unique}.
It is easy to state this property in terms of triangles with common edges,
\ensuremath{\F{triU}\;\mathbin{:}\;\D{Tri}\;\V{\theta}\;\V{\phi}\;\V{\psi}\;\to \;\D{Tri}\;\V{\theta^\prime}\;\V{\phi}\;\V{\psi}\;\to \;\V{\theta}\;\D{=\!\!\!\!=}\;\V{\theta^\prime}},
and then
prove it by induction on the triangles, not edges.
It is thus cheap to obtain \emph{universal properties} in the slices of
$\OPE{}$, asserting the existence of unique morphisms: uniqueness comes for free!

\section{A Proliferation of Functors}

Haskell makes merry with \texttt{class Functor}
and its many subclasses: this scratches but the surface, giving only
\emph{endo}functors from types-and-functions to types-and-functions.
Once we adopt the general notion, functoriality sprouts everywhere, with
the same structures usefully functorial in many ways.

\newcommand{\Id}{\textbf{I}}
\paragraph{Functor.~} A \emph{functor} is a mapping from a source
category $\mathbb{C}$ to a target category $\mathbb{D}$ which
preserves categorical structure.  To specify a structure, we must give
a function $F_o : |\mathbb{C}|\to|\mathbb{D}|$ from source objects
to target objects, together with a family of functions
$F_m : \mathbb{C}(S,T)\to \mathbb{D}(F_o(S),F_o(T))$. The preserved
structure amounts to identity and composition: we must have
that $F_m(\iota_X)=\iota_{F_o(X)}$ and that $F_m(f;g)=F_m(f);F_m(g)$.
Note that there is an identity functor $\Id$ (whose actions on objects
and morphisms are the identity) from $\mathbb{C}$ to itself
and that functors compose (componentwise).

E.g., every \ensuremath{\V{k}\;\mathbin{:}\;\V{K}} induces a functor (\emph{weakening}) from $\OPE{}$ to itself
by scope extension, \ensuremath{(\anonymous \;\C{\mbox{{}-}\!,}\;\V{k})} on objects and \ensuremath{\C{os}} on
morphisms. The very definitions of \ensuremath{\F{oi}} and \ensuremath{\F{\fatsemi}} show that \ensuremath{\C{os}}
preserves \ensuremath{\F{oi}} and \ensuremath{\F{\fatsemi}}.

To see more examples, we need more
categories. Let \ensuremath{\D{Set}}'s objects be types in
Agda's \ensuremath{\D{Set}} universe and $\ensuremath{\D{Set}}(S, T)$ exactly $S\to T$, with the usual identity and
composition. Morphism equality is \emph{pointwise}.
Exercises: make \ensuremath{\D{Bwd}\;\mathbin{:}\;\D{Set}\;\to \;\D{Set}} a functor;
check $(\ensuremath{\D{Lam}},\ensuremath{\F{\uparrow}})$ is a functor from $\OPE{}$ to \ensuremath{\D{Set}}.

Let us plough a different furrow, rich in dependent types, constructing
new categories by \emph{indexing}. If \ensuremath{\V{I}\;\mathbin{:}\;\D{Set}}, we may then take \ensuremath{\V{I}\;\to \;\D{Set}} to be the
category whose objects are \emph{families} of objects in \ensuremath{\D{Set}},
$\ensuremath{\V{S}}, \ensuremath{\V{T}} : \ensuremath{\V{I}\;\to \;\D{Set}}$ with morphisms (implicitly
indexed) families of functions:~ \ensuremath{\V{S}\;\mathrel{\dot{\to}}\;\V{T}\;\mathrel{=}\;\forall\;\{\mskip1.5mu \V{i}\mskip1.5mu\}\;\to \;\V{S}\;\V{i}\;\to \;\V{T}\;\V{i}}.
Morphisms are equal if they map each index to pointwise equal functions.
In the sequel, it will be convenient to abbreviate
\ensuremath{\D{Bwd}\;\V{K}\;\to \;\D{Set}} as \ensuremath{\F{\overline{\black{\V{K}}}}}, for types indexed over scopes.

Dependently typed programming thus offers us a richer seam of
categorical structure than we see in Haskell. This presents an opportunity
to make sense of the categorical taxonomy in terms of concrete
programming examples, and at the same time, organising those programs
and indicating \emph{what to prove}.

\section{Things-with-Thinnings (a Monad)}

Let us acquire the habit of packing terms
together with an
object in the slice of thinnings over their scope,
selecting the
\ensuremath{\F{support}} of the term and discarding unused variables.
Note, \ensuremath{\D{\Uparrow}} is a functor from \ensuremath{\F{\overline{\black{\V{K}}}}} to itself.
\begin{hscode}\SaveRestoreHook
\column{B}{@{}>{\hspre}l<{\hspost}@{}}%
\column{3}{@{}>{\hspre}l<{\hspost}@{}}%
\column{25}{@{}>{\hspre}l<{\hspost}@{}}%
\column{48}{@{}>{\hspre}l<{\hspost}@{}}%
\column{E}{@{}>{\hspre}l<{\hspost}@{}}%
\>[B]{}\mathkw{record}\;\anonymous \D{\Uparrow}\anonymous \;{}\<[25]%
\>[25]{}(\V{T}\;\mathbin{:}\;\F{\overline{\black{\V{K}}}})\;(\V{scope}\;\mathbin{:}\;\D{Bwd}\;\V{K})\;\mathbin{:}\;\D{Set}\;\mathkw{where}\;\mbox{\onelinecomment  \ensuremath{(\V{T}\;\D{\Uparrow}\anonymous )\;\mathbin{:}\;\F{\overline{\black{\V{K}}}}}}{}\<[E]%
\\
\>[B]{}\hsindent{3}{}\<[3]%
\>[3]{}\mathkw{constructor}\;\anonymous \C{\uparrow}\anonymous {}\<[E]%
\\
\>[B]{}\hsindent{3}{}\<[3]%
\>[3]{}\mathkw{field}\;\{\mskip1.5mu \F{support}\mskip1.5mu\}\;\mathbin{:}\;\D{Bwd}\;\V{K};\quad\F{thing}\;\mathbin{:}\;\V{T}\;\F{support};\quad\F{thinning}\;\mathbin{:}\;\F{support}\;\D{\sqsubseteq}\;\V{scope}{}\<[E]%
\\[\blanklineskip]%
\>[B]{}\F{map\!\Uparrow}\;\mathbin{:}\;{}\<[48]%
\>[48]{}(\V{S}\;\mathrel{\dot{\to}}\;\V{T})\;\to \;((\V{S}\;\D{\Uparrow}\anonymous )\;\mathrel{\dot{\to}}\;(\V{T}\;\D{\Uparrow}\anonymous )){}\<[E]%
\\
\>[B]{}\F{map\!\Uparrow}\;\V{f}\;(\V{s}\;\C{\uparrow}\;\V{\theta})\;\mathrel{=}\;\V{f}\;\V{s}\;\C{\uparrow}\;\V{\theta}{}\<[E]%
\ColumnHook
\end{hscode}\resethooks

In fact, the categorical structure of $\OPE{}$ makes \ensuremath{\D{\Uparrow}} a \emph{monad}.
Let us recall the definition.

\paragraph{Monad.~} A functor $M$ from $\mathbb{C}$ to $\mathbb{C}$ gives
rise to a \emph{monad} $(M,\eta,\mu)$ if we can find a pair of natural
transformations, respectively `unit' (`add an $M$ layer') and `multiplication'
(`merge $M$ layers').
\[
\eta_X : \Id(X) \to M(X)
\qquad\qquad
\mu_X : M(M(X)) \to M(X)
\]
subject
to the conditions that merging an added layer yields the identity
(whether the layer added is `outer' or `inner'), and that
adjacent $M$ layers may be merged pairwise in any order.
\[
  \eta_{M(X)};\mu_X = \iota_{M(X)} \qquad\qquad
  M(\eta_X);\mu_X = \iota_{M(X)} \qquad\qquad
  \mu_{M(X)};\mu_X = M(\mu_X);\mu_X
\]

The categorical structure of thinnings makes \ensuremath{\D{\Uparrow}} a monad. Here,
`adding a layer' amounts to `wrapping with a
thinning'. The proof obligations to make $(\ensuremath{\D{\Uparrow}},\ensuremath{\F{unit\!\Uparrow}},\ensuremath{\F{mult\!\Uparrow}})$ a monad are
exactly those required to make $\OPE{}$ a category in the first place.
In particular, things-with-thinnings are easy to thin further, indeed,
parametrically so. In other words, \ensuremath{(\V{T}\;\D{\Uparrow})} is uniformly a functor from
$\OPE{}$ to \ensuremath{\D{Set}}.
\\ \hspace*{ -0.2in}
\begin{tabular}{@{}l|l|l@{}}
\raisebox{0.5in}[0.4in][0in]{\parbox[t]{1.7in}{
\begin{hscode}\SaveRestoreHook
\column{B}{@{}>{\hspre}l<{\hspost}@{}}%
\column{46}{@{}>{\hspre}l<{\hspost}@{}}%
\column{E}{@{}>{\hspre}l<{\hspost}@{}}%
\>[B]{}\F{unit\!\Uparrow}\;\mathbin{:}\;{}\<[46]%
\>[46]{}\V{T}\;\mathrel{\dot{\to}}\;(\V{T}\;\D{\Uparrow}\anonymous ){}\<[E]%
\\
\>[B]{}\F{unit\!\Uparrow}\;\V{t}\;\mathrel{=}\;\V{t}\;\C{\uparrow}\;\F{oi}{}\<[E]%
\ColumnHook
\end{hscode}\resethooks
}}
& \hspace*{ -0.4in}
\raisebox{0.5in}[0.4in][0in]{\parbox[t]{2.5in}{
\begin{hscode}\SaveRestoreHook
\column{B}{@{}>{\hspre}l<{\hspost}@{}}%
\column{46}{@{}>{\hspre}l<{\hspost}@{}}%
\column{E}{@{}>{\hspre}l<{\hspost}@{}}%
\>[B]{}\F{mult\!\Uparrow}\;\mathbin{:}\;{}\<[46]%
\>[46]{}((\V{T}\;\D{\Uparrow}\anonymous )\;\D{\Uparrow}\anonymous )\;\mathrel{\dot{\to}}\;(\V{T}\;\D{\Uparrow}\anonymous ){}\<[E]%
\\
\>[B]{}\F{mult\!\Uparrow}\;((\V{t}\;\C{\uparrow}\;\V{\theta})\;\C{\uparrow}\;\V{\phi})\;\mathrel{=}\;\V{t}\;\C{\uparrow}\;(\V{\theta}\;\F{\fatsemi}\;\V{\phi}){}\<[E]%
\ColumnHook
\end{hscode}\resethooks
}}
& \hspace*{ -0.4in}
\raisebox{0.5in}[0.4in][0in]{\parbox[t]{2.2in}{
\begin{hscode}\SaveRestoreHook
\column{B}{@{}>{\hspre}l<{\hspost}@{}}%
\column{50}{@{}>{\hspre}l<{\hspost}@{}}%
\column{E}{@{}>{\hspre}l<{\hspost}@{}}%
\>[B]{}\F{thin\!\Uparrow}\;\mathbin{:}\;{}\<[50]%
\>[50]{}\V{iz}\;\D{\sqsubseteq}\;\V{jz}\;\to \;\V{T}\;\D{\Uparrow}\;\V{iz}\;\to \;\V{T}\;\D{\Uparrow}\;\V{jz}{}\<[E]%
\\
\>[B]{}\F{thin\!\Uparrow}\;\V{\theta}\;\V{t}\;\mathrel{=}\;\F{mult\!\Uparrow}\;(\V{t}\;\C{\uparrow}\;\V{\theta}){}\<[E]%
\ColumnHook
\end{hscode}\resethooks
}}
\end{tabular}
\\

\newcommand{\Kleisli}[1]{\textbf{Kleisli}(#1)}
%

Shortly, we shall learn how to find the
variables on which a term syntactically depends. However,
merely \emph{allowing} a thinning at the root, \ensuremath{\D{Lam}\;\D{\Uparrow}\;\V{iz}}, yields a
redundant representation, as we may discard
variables at either root or leaves. Let us eliminate redundancy
by \emph{insisting} that a term's
\ensuremath{\F{support}} is \emph{relevant}: a variable retained by the
\ensuremath{\F{thinning}} \emph{must} be used in the \ensuremath{\F{thing}}. Everybody's got to be somewhere.

\section{The Curious Case of the Coproduct in  Slices of $\OPE{}$}

The \ensuremath{\D{\Uparrow}} construction makes crucial use of objects
in the slice category $\OPE{}/\ensuremath{\V{scope}}$, which exhibit useful additional
structure: they are \emph{bit vectors},
with one bit per variable telling whether it has been selected.
Bit vectors inherit Boolean structure, via the
`Naperian' array structure of vectors~\cite{DBLP:conf/esop/Gibbons17}.

\paragraph{Initial object.~} A category $\mathbb{C}$ has initial object
$0$, if there is a unique morphism in $\mathbb{C}(0,X)$ for every $X$.

The \emph{empty type} is famed for this r\^ole for
types-and-functions: empty case analysis gives
the vacuously unique morphism. In $\OPE{}$, the
empty \emph{scope} plays this r\^ole, with the
`constant 0' bit vector as unique morphism.
By return of post, we get $(\ensuremath{\C{[]}},\ensuremath{\F{oe}})$ as the initial object in the
slice category $\OPE{}/\ensuremath{\V{kz}}$. Hence, we can make \emph{constants} with empty support,
i.e., noting that no variable is ($\cdot_R$ for) \emph{relevant}.

\noindent
\begin{tabular}{@{}l|l@{}}
\raisebox{1in}[0.9in][0in]{\parbox[t]{2.9in}{
\begin{hscode}\SaveRestoreHook
\column{B}{@{}>{\hspre}l<{\hspost}@{}}%
\column{21}{@{}>{\hspre}l<{\hspost}@{}}%
\column{26}{@{}>{\hspre}l<{\hspost}@{}}%
\column{30}{@{}>{\hspre}l<{\hspost}@{}}%
\column{47}{@{}>{\hspre}l<{\hspost}@{}}%
\column{E}{@{}>{\hspre}l<{\hspost}@{}}%
\>[B]{}\F{oe}\;\mathbin{:}\;\forall\;{}\<[26]%
\>[26]{}\{\mskip1.5mu \V{kz}\;\mathbin{:}\;\D{Bwd}\;\V{K}\mskip1.5mu\}\;\to \;\C{[]}\;\D{\sqsubseteq}\;\V{kz}{}\<[E]%
\\
\>[B]{}\F{oe}\;\{\mskip1.5mu {}\<[21]%
\>[21]{}\V{iz}\;\C{\mbox{{}-}\!,}\;\V{k}\mskip1.5mu\}\;{}\<[30]%
\>[30]{}\mathrel{=}\;\F{oe}\;\C{o\apo}{}\<[E]%
\\
\>[B]{}\F{oe}\;\{\mskip1.5mu {}\<[21]%
\>[21]{}\C{[]}\mskip1.5mu\}\;{}\<[30]%
\>[30]{}\mathrel{=}\;\C{oz}{}\<[E]%
\\[\blanklineskip]%
\>[B]{}\F{law-}\F{oe}\;\mathbin{:}\;{}\<[47]%
\>[47]{}(\V{\theta}\;\mathbin{:}\;\C{[]}\;\D{\sqsubseteq}\;\V{kz})\;\to \;\V{\theta}\;\D{=\!\!\!\!=}\;\F{oe}{}\<[E]%
\ColumnHook
\end{hscode}\resethooks
}}
&
\raisebox{1in}[0.6in][0in]{\parbox[t]{2.8in}{
\begin{hscode}\SaveRestoreHook
\column{B}{@{}>{\hspre}l<{\hspost}@{}}%
\column{43}{@{}>{\hspre}l<{\hspost}@{}}%
\column{E}{@{}>{\hspre}l<{\hspost}@{}}%
\>[B]{}\F{oe}\F{\slash}\;\mathbin{:}\;{}\<[43]%
\>[43]{}(\V{\theta}\;\mathbin{:}\;\V{iz}\;\D{\sqsubseteq}\;\V{kz})\;\to \;\F{oe}\;\F{\to_{\slash}}\;\V{\theta}{}\<[E]%
\\
\>[B]{}\F{oe}\F{\slash}\;\V{\theta}\;\mathkw{with}\;\F{tri}\;\F{oe}\;\V{\theta}{}\<[E]%
\\
\>[B]{}\V{...}\;\mid \;\V{t}\;\mathkw{rewrite}\;\F{law-}\F{oe}\;(\F{oe}\;\F{\fatsemi}\;\V{\theta})\;\mathrel{=}\;\F{oe}\;\C{,}\;\V{t}{}\<[E]%
\ColumnHook
\end{hscode}\resethooks
}}
\\
\raisebox{0.3in}[0.2in][0in]{\parbox[t]{2.9in}{
\begin{hscode}\SaveRestoreHook
\column{B}{@{}>{\hspre}l<{\hspost}@{}}%
\column{25}{@{}>{\hspre}l<{\hspost}@{}}%
\column{42}{@{}>{\hspre}l<{\hspost}@{}}%
\column{E}{@{}>{\hspre}l<{\hspost}@{}}%
\>[B]{}\mathkw{data}\;\D{One}_{R}\;{}\<[25]%
\>[25]{}\mathbin{:}\;\F{\overline{\black{\V{K}}}}\;\mathkw{where}\;{}\<[42]%
\>[42]{}\C{\langle\rangle}\;\mathbin{:}\;\D{One}_{R}\;\C{[]}{}\<[E]%
\ColumnHook
\end{hscode}\resethooks
}}
&
\raisebox{0.3in}[0.2in][0in]{\parbox[t]{2.8in}{
\begin{hscode}\SaveRestoreHook
\column{B}{@{}>{\hspre}l<{\hspost}@{}}%
\column{43}{@{}>{\hspre}l<{\hspost}@{}}%
\column{E}{@{}>{\hspre}l<{\hspost}@{}}%
\>[B]{}\F{\langle\rangle}_{R}\;\mathbin{:}\;{}\<[43]%
\>[43]{}\D{One}_{R}\;\D{\Uparrow}\;\V{kz};\quad\F{\langle\rangle}_{R}\;\mathrel{=}\;\C{\langle\rangle}\;\C{\uparrow}\;\F{oe}{}\<[E]%
\ColumnHook
\end{hscode}\resethooks
}}
\end{tabular}
\vspace*{0.05in}



We should expect the constant to be the trivial case of some notion
of \emph{relevant pairing}, induced by \emph{coproducts} in the slice category.
If we have two objects in $\OPE{}/\ensuremath{\V{kz}}$ representing two subscopes, \ensuremath{(\V{iz,}\;\V{\theta})} and
\ensuremath{(\V{jz,}\;\V{\phi})}, there should
be a smallest subscope which includes both: pairwise disjunction
of bit vectors.

\paragraph{Coproduct.~} Objects $S$ and $T$ of category $\mathbb{C}$ have
a coproduct object $S + T$ if there are morphisms $l\in \mathbb{C}(S, S + T)$ and
$r\in \mathbb{C}(T, S + T)$ such that every pair
$f\in \mathbb{C}(S, U)$ and $g\in \mathbb{C}(T, U)$ factors
through a unique $h\in \mathbb{C}(S + T, U)$ so that
$f = l;h$ and $g = r;h$. In \ensuremath{\D{Set}}, we may take $S + T$ to be
the \emph{disjoint union} of $S$ and $T$, with $l$ and $r$ its injections
and $h$ the \emph{case analysis} whose branches are $f$ and $g$.

However, we are not working in \ensuremath{\D{Set}}, but in a slice category.
Any category theorist will tell you that slice categories $\mathbb{C}/I$
inherit \emph{colimit} structure (characterized by universal out-arrows)
from $\mathbb{C}$, as indeed we just saw with the initial object. If $\OPE{}$
has coproducts, too, we are done! Taking \ensuremath{\V{K}\;\mathrel{=}\;\D{One}},
let us seek the coproduct of two singletons, $S = T = \ensuremath{\C{[]}\;\C{\mbox{{}-}\!,}\;\C{\langle\rangle}}$.
Construct one diagram by taking $U = \ensuremath{\C{[]}\;\C{\mbox{{}-}\!,}\;\C{\langle\rangle}}$ and $f = g = \ensuremath{\F{oi}}$,
ensuring that our only candidate for $S + T$ is again the
singleton \ensuremath{\C{[]}\;\C{\mbox{{}-}\!,}\;\C{\langle\rangle}}, with $l = r = \ensuremath{\F{oi}}$, making $h = \ensuremath{\F{oi}}$.
Nothing else can sit between $S, T$ and $U$.
\newcommand{\stackb}[2]{\mbox{$\begin{array}[b]{@{}c@{}}#1 \\ #2\end{array}$}}
\newcommand{\stackt}[2]{\mbox{$\begin{array}[t]{@{}c@{}}#1 \\ #2\end{array}$}}
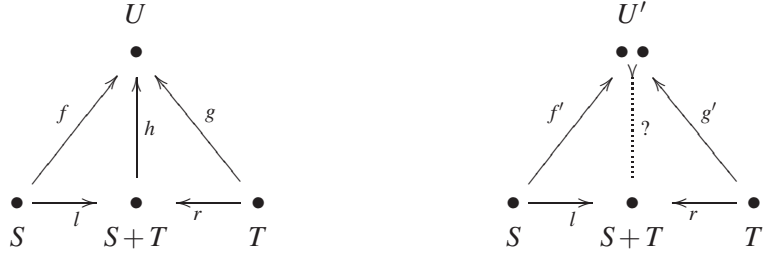
\[\xymatrix{
        & \stackb{U}{\bullet} &         &&&   & \stackb{U'}{\bullet\;\bullet} & \\
\stackt{\bullet}{S} \ar[ur]^f \ar[r]_l & \stackt{\bullet}{S+T} \ar[u]_h & \stackt{\bullet}{T} \ar[l]^r \ar[ul]_g &&&
\stackt{\bullet}{S} \ar[ur]^{f'} \ar[r]_l & \stackt{\bullet}{S+T} \ar@{..<}[u]_{?} & \stackt{\bullet}{T} \ar[l]^r \ar[ul]_{g'}
}\]
Now begin a different diagram, with $U' = \ensuremath{\C{[]}\;\C{\mbox{{}-}\!,}\;\C{\langle\rangle}\;\C{\mbox{{}-}\!,}\;\C{\langle\rangle}}$,
allowing $f' = \ensuremath{\C{oz}\;\C{os}\;\C{o\apo}}$ and $g' = \ensuremath{\C{oz}\;\C{o\apo}\;\C{os}}$. No $h'$
post-composes $l$ and $r$ (both \ensuremath{\F{oi}}, making $h'$ itself)
to yield $f'$ and $g'$ respectively. We do not get coproducts.

Fortunately, we get what we need: $\OPE{}$ may not have coproducts, but its
\emph{slices} do.
Examine the data: two subscopes of some \ensuremath{\V{kz}}, \ensuremath{\V{\theta}\;\mathbin{:}\;\V{iz}\;\D{\sqsubseteq}\;\V{kz}} and \ensuremath{\V{\phi}\;\mathbin{:}\;\V{jz}\;\D{\sqsubseteq}\;\V{kz}}. Their
coproduct must be some \ensuremath{\V{\psi}\;\mathbin{:}\;\V{ijz}\;\D{\sqsubseteq}\;\V{kz}}, where our $l$ and $r$ must be
triangles \ensuremath{\D{Tri}\;\V{\theta^\prime}\;\V{\psi}\;\V{\theta}} and \ensuremath{\D{Tri}\;\V{\phi^\prime}\;\V{\psi}\;\V{\phi}}, giving morphisms in
\ensuremath{\V{\theta}\;\F{\to_{\slash}}\;\V{\psi}} and \ensuremath{\V{\phi}\;\F{\to_{\slash}}\;\V{\psi}}. Choose \ensuremath{\V{\psi}}
to be pointwise disjunction of \ensuremath{\V{\theta}} and \ensuremath{\V{\phi}}, minimizing \ensuremath{\V{ijz}}: \ensuremath{\V{\theta^\prime}} and \ensuremath{\V{\phi^\prime}} will then \emph{cover} \ensuremath{\V{ijz}}.
\begin{hscode}\SaveRestoreHook
\column{B}{@{}>{\hspre}l<{\hspost}@{}}%
\column{3}{@{}>{\hspre}l<{\hspost}@{}}%
\column{9}{@{}>{\hspre}l<{\hspost}@{}}%
\column{25}{@{}>{\hspre}l<{\hspost}@{}}%
\column{71}{@{}>{\hspre}l<{\hspost}@{}}%
\column{76}{@{}>{\hspre}l<{\hspost}@{}}%
\column{95}{@{}>{\hspre}l<{\hspost}@{}}%
\column{114}{@{}>{\hspre}l<{\hspost}@{}}%
\column{148}{@{}>{\hspre}l<{\hspost}@{}}%
\column{153}{@{}>{\hspre}l<{\hspost}@{}}%
\column{158}{@{}>{\hspre}l<{\hspost}@{}}%
\column{163}{@{}>{\hspre}l<{\hspost}@{}}%
\column{E}{@{}>{\hspre}l<{\hspost}@{}}%
\>[B]{}\mathkw{data}\;\D{Cover}\;{}\<[25]%
\>[25]{}(\V{ov}\;\mathbin{:}\;\D{Two})\;\mathbin{:}\;{}\<[71]%
\>[71]{}\V{iz}\;\D{\sqsubseteq}\;\V{ijz}\;\to \;\V{jz}\;\D{\sqsubseteq}\;\V{ijz}\;\to \;\D{Set}\;\mathkw{where}{}\<[E]%
\\
\>[B]{}\hsindent{3}{}\<[3]%
\>[3]{}\anonymous \C{c\apo s}\;{}\<[9]%
\>[9]{}\mathbin{:}\;{}\<[95]%
\>[95]{}\D{Cover}\;\V{ov}\;\V{\theta}\;\V{\phi}\;\to \;{}\<[114]%
\>[114]{}\D{Cover}\;\V{ov}\;{}\<[148]%
\>[148]{}(\V{\theta}\;{}\<[153]%
\>[153]{}\C{o\apo})\;{}\<[158]%
\>[158]{}(\V{\phi}\;{}\<[163]%
\>[163]{}\C{os}){}\<[E]%
\\
\>[B]{}\hsindent{3}{}\<[3]%
\>[3]{}\anonymous \C{cs\apo}\;{}\<[9]%
\>[9]{}\mathbin{:}\;{}\<[95]%
\>[95]{}\D{Cover}\;\V{ov}\;\V{\theta}\;\V{\phi}\;\to \;{}\<[114]%
\>[114]{}\D{Cover}\;\V{ov}\;{}\<[148]%
\>[148]{}(\V{\theta}\;{}\<[153]%
\>[153]{}\C{os})\;{}\<[158]%
\>[158]{}(\V{\phi}\;{}\<[163]%
\>[163]{}\C{o\apo}){}\<[E]%
\\
\>[B]{}\hsindent{3}{}\<[3]%
\>[3]{}\anonymous \C{css}\;{}\<[9]%
\>[9]{}\mathbin{:}\;{}\<[76]%
\>[76]{}\{\mskip1.5mu \V{both}\;\mathbin{:}\;\F{T\!t}\;\V{ov}\mskip1.5mu\}\;\to \;{}\<[95]%
\>[95]{}\D{Cover}\;\V{ov}\;\V{\theta}\;\V{\phi}\;\to \;{}\<[114]%
\>[114]{}\D{Cover}\;\V{ov}\;{}\<[148]%
\>[148]{}(\V{\theta}\;{}\<[153]%
\>[153]{}\C{os})\;{}\<[158]%
\>[158]{}(\V{\phi}\;{}\<[163]%
\>[163]{}\C{os}){}\<[E]%
\\
\>[B]{}\hsindent{3}{}\<[3]%
\>[3]{}\C{czz}\;{}\<[9]%
\>[9]{}\mathbin{:}\;{}\<[114]%
\>[114]{}\D{Cover}\;\V{ov}\;{}\<[153]%
\>[153]{}\C{oz}\;{}\<[163]%
\>[163]{}\C{oz}{}\<[E]%
\ColumnHook
\end{hscode}\resethooks
The flag, \ensuremath{\V{ov}},
determines whether \emph{overlap} is permitted: with \ensuremath{\C{t\!t}} for
coproducts and \ensuremath{\C{f\!f}} for \emph{partitions}.
No constructor allows both \ensuremath{\V{\theta}} and \ensuremath{\V{\phi}} to
omit a target variable, so everybody's got to be somewhere.
Let us compute
the coproduct, \ensuremath{\V{\psi}} then check that any other diagram for
some \ensuremath{\V{\psi^\prime}} yields a \ensuremath{\V{\psi}\;\F{\to_{\slash}}\;\V{\psi^\prime}}.
\vspace*{ -0.2in} \\
\parbox{4in}{
\begin{hscode}\SaveRestoreHook
\column{B}{@{}>{\hspre}l<{\hspost}@{}}%
\column{7}{@{}>{\hspre}l<{\hspost}@{}}%
\column{50}{@{}>{\hspre}l<{\hspost}@{}}%
\column{75}{@{}>{\hspre}l<{\hspost}@{}}%
\column{136}{@{}>{\hspre}l<{\hspost}@{}}%
\column{E}{@{}>{\hspre}l<{\hspost}@{}}%
\>[B]{}\F{cop}\;{}\<[7]%
\>[7]{}\mathbin{:}\;{}\<[50]%
\>[50]{}(\V{\theta}\;\mathbin{:}\;\V{iz}\;\D{\sqsubseteq}\;\V{kz})\;(\V{\phi}\;\mathbin{:}\;\V{jz}\;\D{\sqsubseteq}\;\V{kz})\;\to \;{}\<[E]%
\\[\blanklineskip]%
\>[50]{}\D{\Upsigma}\;\anonymous \;\lambda \;\V{ijz}\;\to \;{}\<[75]%
\>[75]{}\D{\Upsigma}\;(\V{ijz}\;\D{\sqsubseteq}\;\V{kz})\;\lambda \;\V{\psi}\;\to \;{}\<[E]%
\\
\>[50]{}\D{\Upsigma}\;(\V{iz}\;\D{\sqsubseteq}\;\V{ijz})\;\lambda \;\V{\theta^\prime}\;\to \;{}\<[75]%
\>[75]{}\D{\Upsigma}\;(\V{jz}\;\D{\sqsubseteq}\;\V{ijz})\;\lambda \;\V{\phi^\prime}\;\to \;{}\<[E]%
\\
\>[50]{}\D{Tri}\;\V{\theta^\prime}\;\V{\psi}\;\V{\theta}\;\F{\times}\;\D{Cover}\;\C{t\!t}\;\V{\theta^\prime}\;\V{\phi^\prime}\;\F{\times}\;\D{Tri}\;\V{\phi^\prime}\;\V{\psi}\;\V{\phi}{}\<[E]%
\\[\blanklineskip]%
\>[B]{}\F{copU}\;{}\<[7]%
\>[7]{}\mathbin{:}\;{}\<[136]%
\>[136]{}\D{Tri}\;\V{\theta^\prime}\;\V{\psi}\;\V{\theta}\;\to \;\D{Cover}\;\C{t\!t}\;\V{\theta^\prime}\;\V{\phi^\prime}\;\to \;\D{Tri}\;\V{\phi^\prime}\;\V{\psi}\;\V{\phi}\;\to \;{}\<[E]%
\\
\>[136]{}\V{\theta}\;\F{\to_{\slash}}\;\V{\psi^\prime}\;\to \;\V{\phi}\;\F{\to_{\slash}}\;\V{\psi^\prime}\;\to \;\V{\psi}\;\F{\to_{\slash}}\;\V{\psi^\prime}{}\<[E]%
\ColumnHook
\end{hscode}\resethooks
}
\parbox{2in}{
\[\xymatrix{
 \ensuremath{\V{iz}} \ar[rrrd]^{\ensuremath{\V{\theta}}} \ar@{ -->}[rd]_{\ensuremath{\V{\theta^\prime}}} &&& \\
 &  \!\!\!\!\langle\;\ensuremath{\V{ijz}} \ar@{ -->}[rr]^{\ensuremath{\V{\psi}}\qquad} && \ensuremath{\V{kz}} \\ 
 \ensuremath{\V{jz}} \ar[rrru]_{\ensuremath{\V{\phi}}} \ar@{ -->}[ru]^{\ensuremath{\V{\phi^\prime}}} &&& \\
}\]}
\\
where the $\langle$ in the diagram indicates that the two incoming arrows form a \ensuremath{\D{Cover}}.

The recursive steps in \ensuremath{\F{cop}}'s implementation
work explicitly with the two-dimensional triangles and coverings,
using \ensuremath{\C{!}} to hide their boundaries (thinnings) and their boundaries' boundaries (scopes).
\newpage   
\begin{hscode}\SaveRestoreHook
\column{B}{@{}>{\hspre}l<{\hspost}@{}}%
\column{10}{@{}>{\hspre}l<{\hspost}@{}}%
\column{19}{@{}>{\hspre}l<{\hspost}@{}}%
\column{24}{@{}>{\hspre}l<{\hspost}@{}}%
\column{66}{@{}>{\hspre}l<{\hspost}@{}}%
\column{78}{@{}>{\hspre}l<{\hspost}@{}}%
\column{84}{@{}>{\hspre}l<{\hspost}@{}}%
\column{89}{@{}>{\hspre}l<{\hspost}@{}}%
\column{94}{@{}>{\hspre}l<{\hspost}@{}}%
\column{100}{@{}>{\hspre}l<{\hspost}@{}}%
\column{E}{@{}>{\hspre}l<{\hspost}@{}}%
\>[B]{}\F{cop}\;(\V{\theta}\;{}\<[10]%
\>[10]{}\C{o\apo})\;(\V{\phi}\;{}\<[19]%
\>[19]{}\C{o\apo})\;{}\<[24]%
\>[24]{}\mathrel{=}\;\mathkw{let}\;\C{!}\;\C{!}\;\C{!}\;\C{!}\;\V{tl}\;\C{,}\;\V{c}\;\C{,}\;\V{tr}\;\mathrel{=}\;\F{cop}\;\V{\theta}\;\V{\phi}\;\mathkw{in}\;{}\<[66]%
\>[66]{}\C{!}\;\C{!}\;\C{!}\;\C{!}\;\V{tl}\;{}\<[78]%
\>[78]{}\C{t\mbox{{}-}\apo\!\apo}\;{}\<[84]%
\>[84]{}\C{,}\;\V{c}\;{}\<[94]%
\>[94]{}\C{,}\;\V{tr}\;{}\<[100]%
\>[100]{}\C{t\mbox{{}-}\apo\!\apo}{}\<[E]%
\\
\>[B]{}\F{cop}\;(\V{\theta}\;{}\<[10]%
\>[10]{}\C{o\apo})\;(\V{\phi}\;{}\<[19]%
\>[19]{}\C{os})\;{}\<[24]%
\>[24]{}\mathrel{=}\;\mathkw{let}\;\C{!}\;\C{!}\;\C{!}\;\C{!}\;\V{tl}\;\C{,}\;\V{c}\;\C{,}\;\V{tr}\;\mathrel{=}\;\F{cop}\;\V{\theta}\;\V{\phi}\;\mathkw{in}\;{}\<[66]%
\>[66]{}\C{!}\;\C{!}\;\C{!}\;\C{!}\;\V{tl}\;{}\<[78]%
\>[78]{}\C{t\apo\!s\apo}\;{}\<[84]%
\>[84]{}\C{,}\;\V{c}\;{}\<[89]%
\>[89]{}\C{c\apo s}\;{}\<[94]%
\>[94]{}\C{,}\;\V{tr}\;{}\<[100]%
\>[100]{}\C{tsss}{}\<[E]%
\\
\>[B]{}\F{cop}\;(\V{\theta}\;{}\<[10]%
\>[10]{}\C{os})\;(\V{\phi}\;{}\<[19]%
\>[19]{}\C{o\apo})\;{}\<[24]%
\>[24]{}\mathrel{=}\;\mathkw{let}\;\C{!}\;\C{!}\;\C{!}\;\C{!}\;\V{tl}\;\C{,}\;\V{c}\;\C{,}\;\V{tr}\;\mathrel{=}\;\F{cop}\;\V{\theta}\;\V{\phi}\;\mathkw{in}\;{}\<[66]%
\>[66]{}\C{!}\;\C{!}\;\C{!}\;\C{!}\;\V{tl}\;{}\<[78]%
\>[78]{}\C{tsss}\;{}\<[84]%
\>[84]{}\C{,}\;\V{c}\;{}\<[89]%
\>[89]{}\C{cs\apo}\;{}\<[94]%
\>[94]{}\C{,}\;\V{tr}\;{}\<[100]%
\>[100]{}\C{t\apo\!s\apo}{}\<[E]%
\\
\>[B]{}\F{cop}\;(\V{\theta}\;{}\<[10]%
\>[10]{}\C{os})\;(\V{\phi}\;{}\<[19]%
\>[19]{}\C{os})\;{}\<[24]%
\>[24]{}\mathrel{=}\;\mathkw{let}\;\C{!}\;\C{!}\;\C{!}\;\C{!}\;\V{tl}\;\C{,}\;\V{c}\;\C{,}\;\V{tr}\;\mathrel{=}\;\F{cop}\;\V{\theta}\;\V{\phi}\;\mathkw{in}\;{}\<[66]%
\>[66]{}\C{!}\;\C{!}\;\C{!}\;\C{!}\;\V{tl}\;{}\<[78]%
\>[78]{}\C{tsss}\;{}\<[84]%
\>[84]{}\C{,}\;\V{c}\;{}\<[89]%
\>[89]{}\C{css}\;{}\<[94]%
\>[94]{}\C{,}\;\V{tr}\;{}\<[100]%
\>[100]{}\C{tsss}{}\<[E]%
\\
\>[B]{}\F{cop}\;{}\<[10]%
\>[10]{}\C{oz}\;{}\<[19]%
\>[19]{}\C{oz}\;{}\<[24]%
\>[24]{}\mathrel{=}\;{}\<[66]%
\>[66]{}\C{!}\;\C{!}\;\C{!}\;\C{!}\;{}\<[78]%
\>[78]{}\C{tzzz}\;{}\<[84]%
\>[84]{}\C{,}\;{}\<[89]%
\>[89]{}\C{czz}\;{}\<[94]%
\>[94]{}\C{,}\;{}\<[100]%
\>[100]{}\C{tzzz}{}\<[E]%
\ColumnHook
\end{hscode}\resethooks
The \ensuremath{\F{copU}} proof goes by induction on the triangles which share \ensuremath{\V{\psi^\prime}}
and inversion of the coproduct.

\noindent
\begin{tabular}{@{}ll@{}}
\parbox{4.4in}{
A further useful property of coproduct diagrams is that we can selectively refine
them by a thinning into the covered scope.
\begin{hscode}\SaveRestoreHook
\column{B}{@{}>{\hspre}l<{\hspost}@{}}%
\column{97}{@{}>{\hspre}l<{\hspost}@{}}%
\column{E}{@{}>{\hspre}l<{\hspost}@{}}%
\>[B]{}\F{subCop}\;\mathbin{:}\;{}\<[97]%
\>[97]{}(\V{\psi}\;\mathbin{:}\;\V{kz}\;\D{\sqsubseteq}\;\V{kz'})\;\to \;\D{Cover}\;\V{ov}\;\V{\theta^\prime}\;\V{\phi^\prime}\;\to \;{}\<[E]%
\\
\>[97]{}\D{\Upsigma}\;\anonymous \;\lambda \;\V{iz}\;\to \;\D{\Upsigma}\;\anonymous \;\lambda \;\V{jz}\;\to \;\D{\Upsigma}\;(\V{iz}\;\D{\sqsubseteq}\;\V{kz})\;\lambda \;\V{\theta}\;\to \;\D{\Upsigma}\;(\V{jz}\;\D{\sqsubseteq}\;\V{kz})\;\lambda \;\V{\phi}\;\to \;{}\<[E]%
\\
\>[97]{}\D{\Upsigma}\;(\V{iz}\;\D{\sqsubseteq}\;\V{iz'})\;\lambda \;\V{\psi_0}\;\to \;\D{\Upsigma}\;(\V{jz}\;\D{\sqsubseteq}\;\V{jz'})\;\lambda \;\V{\psi_1}\;\to \;\D{Cover}\;\V{ov}\;\V{\theta}\;\V{\phi}{}\<[E]%
\ColumnHook
\end{hscode}\resethooks
The implementation is a straightforward induction on the diagram.
}
&
\parbox{2in}{
\[\xymatrix{
\ensuremath{\V{iz}} \ar@{ -->}[r]^{\ensuremath{\V{\psi_0}}} \ar@{ -->}[dr]^{\ensuremath{\V{\theta}}}
   & \ensuremath{\V{iz'}} \ar[dr]^{\ensuremath{\V{\theta^\prime}}}                  & \\
   & \!\!\!\!\langle\;\ensuremath{\V{kz}} \ar[r]^{\ensuremath{\V{\psi}}}    & \!\!\!\!\langle\;\ensuremath{\V{kz'}} \\
\ensuremath{\V{jz}} \ar@{ -->}[r]_{\ensuremath{\V{\psi_1}}} \ar@{ -->}[ur]_{\ensuremath{\V{\phi}}}
   & \ensuremath{\V{jz'}} \ar[ur]_{\ensuremath{\V{\phi^\prime}}}                  & \\
}\]}
\end{tabular}

The payoff from coproducts is the type of \emph{relevant pairs} ---
the co-de-Bruijn touchstone:
\\

\noindent
\hspace*{ -0.3in}
\raisebox{0.9in}[0.8in][0in]{
\parbox[t]{3.8in}{
\begin{hscode}\SaveRestoreHook
\column{B}{@{}>{\hspre}l<{\hspost}@{}}%
\column{3}{@{}>{\hspre}l<{\hspost}@{}}%
\column{10}{@{}>{\hspre}l<{\hspost}@{}}%
\column{17}{@{}>{\hspre}l<{\hspost}@{}}%
\column{26}{@{}>{\hspre}l<{\hspost}@{}}%
\column{E}{@{}>{\hspre}l<{\hspost}@{}}%
\>[B]{}\mathkw{record}\;\anonymous \D{\times}_R\anonymous \;{}\<[26]%
\>[26]{}(\V{S}\;\V{T}\;\mathbin{:}\;\F{\overline{\black{\V{K}}}})\;(\V{ijz}\;\mathbin{:}\;\D{Bwd}\;\V{K})\;\mathbin{:}\;\D{Set}\;\mathkw{where}{}\<[E]%
\\
\>[B]{}\hsindent{3}{}\<[3]%
\>[3]{}\mathkw{constructor}\;\C{pair}{}\<[E]%
\\
\>[B]{}\hsindent{3}{}\<[3]%
\>[3]{}\mathkw{field}\;{}\<[10]%
\>[10]{}\F{outl}\;{}\<[17]%
\>[17]{}\mathbin{:}\;\V{S}\;\D{\Uparrow}\;\V{ijz};\quad\F{outr}\;\mathbin{:}\;\V{T}\;\D{\Uparrow}\;\V{ijz}{}\<[E]%
\\
\>[10]{}\F{cover}\;{}\<[17]%
\>[17]{}\mathbin{:}\;\D{Cover}\;\C{t\!t}\;(\F{thinning}\;\F{outl})\;(\F{thinning}\;\F{outr}){}\<[E]%
\ColumnHook
\end{hscode}\resethooks
}}
\vrule \hspace*{ -0.4in}
\raisebox{0.9in}[0.8in][0in]{\parbox[t]{2in}{
\begin{hscode}\SaveRestoreHook
\column{B}{@{}>{\hspre}l<{\hspost}@{}}%
\column{3}{@{}>{\hspre}l<{\hspost}@{}}%
\column{8}{@{}>{\hspre}l<{\hspost}@{}}%
\column{51}{@{}>{\hspre}l<{\hspost}@{}}%
\column{E}{@{}>{\hspre}l<{\hspost}@{}}%
\>[B]{}\anonymous \F{,}_R\anonymous \;\mathbin{:}\;{}\<[51]%
\>[51]{}\V{S}\;\D{\Uparrow}\;\V{kz}\;\to \;\V{T}\;\D{\Uparrow}\;\V{kz}\;\to \;(\V{S}\;\D{\times}_R\;\V{T})\;\D{\Uparrow}\;\V{kz}{}\<[E]%
\\
\>[B]{}(\V{s}\;\C{\uparrow}\;\V{\theta})\;\F{,}_R\;(\V{t}\;\C{\uparrow}\;\V{\phi})\;\mathrel{=}{}\<[E]%
\\
\>[B]{}\hsindent{3}{}\<[3]%
\>[3]{}\mathkw{let}\;{}\<[8]%
\>[8]{}\C{!}\;\V{\psi}\;\C{,}\;\V{\theta^\prime}\;\C{,}\;\V{\phi^\prime}\;\C{,}\;\anonymous \;\C{,}\;\V{c}\;\C{,}\;\anonymous \;\mathrel{=}\;\F{cop}\;\V{\theta}\;\V{\phi}{}\<[E]%
\\
\>[B]{}\hsindent{3}{}\<[3]%
\>[3]{}\mathkw{in}\;{}\<[8]%
\>[8]{}\C{pair}\;(\V{s}\;\C{\uparrow}\;\V{\theta^\prime})\;(\V{t}\;\C{\uparrow}\;\V{\phi^\prime})\;\V{c}\;\C{\uparrow}\;\V{\psi}{}\<[E]%
\ColumnHook
\end{hscode}\resethooks
}}\\
The corresponding projections are readily definable.

\noindent
\raisebox{0.5in}[0.4in][0in]{\parbox[t]{3in}{
\begin{hscode}\SaveRestoreHook
\column{B}{@{}>{\hspre}l<{\hspost}@{}}%
\column{53}{@{}>{\hspre}l<{\hspost}@{}}%
\column{E}{@{}>{\hspre}l<{\hspost}@{}}%
\>[B]{}\F{outl}_R\;\mathbin{:}\;{}\<[53]%
\>[53]{}(\V{S}\;\D{\times}_R\;\V{T})\;\D{\Uparrow}\;\V{kz}\;\to \;\V{S}\;\D{\Uparrow}\;\V{kz}{}\<[E]%
\\
\>[B]{}\F{outl}_R\;(\C{pair}\;\V{s}\;\anonymous \;\anonymous \;\C{\uparrow}\;\V{\psi})\;\mathrel{=}\;\F{thin\!\Uparrow}\;\V{\psi}\;\V{s}{}\<[E]%
\ColumnHook
\end{hscode}\resethooks
}}
\vrule
\raisebox{0.5in}[0.4in][0in]{\parbox[t]{2.6in}{
\begin{hscode}\SaveRestoreHook
\column{B}{@{}>{\hspre}l<{\hspost}@{}}%
\column{53}{@{}>{\hspre}l<{\hspost}@{}}%
\column{E}{@{}>{\hspre}l<{\hspost}@{}}%
\>[B]{}\F{outr}_R\;\mathbin{:}\;{}\<[53]%
\>[53]{}(\V{S}\;\D{\times}_R\;\V{T})\;\D{\Uparrow}\;\V{kz}\;\to \;\V{T}\;\D{\Uparrow}\;\V{kz}{}\<[E]%
\\
\>[B]{}\F{outr}_R\;(\C{pair}\;\anonymous \;\V{t}\;\anonymous \;\C{\uparrow}\;\V{\psi})\;\mathrel{=}\;\F{thin\!\Uparrow}\;\V{\psi}\;\V{t}{}\<[E]%
\ColumnHook
\end{hscode}\resethooks
}}

\section{Monoidal Structure of Order-Preserving Embeddings}

Variable bindings extend scopes. The \ensuremath{\C{\uplambda}} construct does just one
`snoc', but binding can be simultaneous, so the monoidal
structure on $\OPE{}$ induced by concatenation is what we need.
\\
\begin{tabular}{@{}l|l@{}}
\hspace*{ -0.3in}
\raisebox{0.9in}[0.8in][0in]{\parbox[t]{2.8in}{
\begin{hscode}\SaveRestoreHook
\column{B}{@{}>{\hspre}l<{\hspost}@{}}%
\column{17}{@{}>{\hspre}l<{\hspost}@{}}%
\column{32}{@{}>{\hspre}l<{\hspost}@{}}%
\column{E}{@{}>{\hspre}l<{\hspost}@{}}%
\>[B]{}\anonymous \F{+\!\!+}\anonymous \;\mathbin{:}\;{}\<[32]%
\>[32]{}\D{Bwd}\;\V{K}\;\to \;\D{Bwd}\;\V{K}\;\to \;\D{Bwd}\;\V{K}{}\<[E]%
\\
\>[B]{}\V{kz}\;\F{+\!\!+}\;\C{[]}\;{}\<[17]%
\>[17]{}\mathrel{=}\;\V{kz}{}\<[E]%
\\
\>[B]{}\V{kz}\;\F{+\!\!+}\;(\V{iz}\;\C{\mbox{{}-}\!,}\;\V{j})\;{}\<[17]%
\>[17]{}\mathrel{=}\;(\V{kz}\;\F{+\!\!+}\;\V{iz})\;\C{\mbox{{}-}\!,}\;\V{j}{}\<[E]%
\ColumnHook
\end{hscode}\resethooks
}}
& \hspace*{ -0.4in}
\raisebox{0.9in}[0.8in][0in]{\parbox[t]{3in}{
\begin{hscode}\SaveRestoreHook
\column{B}{@{}>{\hspre}l<{\hspost}@{}}%
\column{14}{@{}>{\hspre}l<{\hspost}@{}}%
\column{19}{@{}>{\hspre}l<{\hspost}@{}}%
\column{24}{@{}>{\hspre}l<{\hspost}@{}}%
\column{57}{@{}>{\hspre}l<{\hspost}@{}}%
\column{E}{@{}>{\hspre}l<{\hspost}@{}}%
\>[B]{}\anonymous \F{+\!\!+}_{\D{\sqsubseteq}}\anonymous \;\mathbin{:}\;{}\<[57]%
\>[57]{}\V{iz}\;\D{\sqsubseteq}\;\V{jz}\;\to \;\V{iz'}\;\D{\sqsubseteq}\;\V{jz'}\;\to \;(\V{iz}\;\F{+\!\!+}\;\V{iz'})\;\D{\sqsubseteq}\;(\V{jz}\;\F{+\!\!+}\;\V{jz'}){}\<[E]%
\\
\>[B]{}\V{\theta}\;\F{+\!\!+}_{\D{\sqsubseteq}}\;{}\<[14]%
\>[14]{}\C{oz}\;{}\<[19]%
\>[19]{}\mathrel{=}\;{}\<[24]%
\>[24]{}\V{\theta}{}\<[E]%
\\
\>[B]{}\V{\theta}\;\F{+\!\!+}_{\D{\sqsubseteq}}\;(\V{\phi}\;{}\<[14]%
\>[14]{}\C{os})\;{}\<[19]%
\>[19]{}\mathrel{=}\;({}\<[24]%
\>[24]{}\V{\theta}\;\F{+\!\!+}_{\D{\sqsubseteq}}\;\V{\phi})\;\C{os}{}\<[E]%
\\
\>[B]{}\V{\theta}\;\F{+\!\!+}_{\D{\sqsubseteq}}\;(\V{\phi}\;{}\<[14]%
\>[14]{}\C{o\apo})\;{}\<[19]%
\>[19]{}\mathrel{=}\;({}\<[24]%
\>[24]{}\V{\theta}\;\F{+\!\!+}_{\D{\sqsubseteq}}\;\V{\phi})\;\C{o\apo}{}\<[E]%
\ColumnHook
\end{hscode}\resethooks
}}
\end{tabular}

Concatenation further extends to \ensuremath{\D{Cover}}ings, allowing us to build them in chunks.
\begin{hscode}\SaveRestoreHook
\column{B}{@{}>{\hspre}l<{\hspost}@{}}%
\column{28}{@{}>{\hspre}l<{\hspost}@{}}%
\column{142}{@{}>{\hspre}l<{\hspost}@{}}%
\column{E}{@{}>{\hspre}l<{\hspost}@{}}%
\>[B]{}\anonymous \F{+\!\!+}_C\anonymous \;\mathbin{:}\;{}\<[142]%
\>[142]{}\D{Cover}\;\V{ov}\;\V{\theta}\;\V{\phi}\;\to \;\D{Cover}\;\V{ov}\;\V{\theta^\prime}\;\V{\phi^\prime}\;\to \;\D{Cover}\;\V{ov}\;(\V{\theta}\;\F{+\!\!+}_{\D{\sqsubseteq}}\;\V{\theta^\prime})\;(\V{\phi}\;\F{+\!\!+}_{\D{\sqsubseteq}}\;\V{\phi^\prime}){}\<[E]%
\\
\>[B]{}\V{c}\;\F{+\!\!+}_C\;(\V{d}\;\C{c\apo s})\;{}\<[28]%
\>[28]{}\mathrel{=}\;(\V{c}\;\F{+\!\!+}_C\;\V{d})\;\C{c\apo s}{}\<[E]%
\\
\>[B]{}\V{c}\;\F{+\!\!+}_C\;(\V{d}\;\C{cs\apo})\;{}\<[28]%
\>[28]{}\mathrel{=}\;(\V{c}\;\F{+\!\!+}_C\;\V{d})\;\C{cs\apo}{}\<[E]%
\\
\>[B]{}\V{c}\;\F{+\!\!+}_C\;(\anonymous \C{css}\;\{\mskip1.5mu \V{both}\;\mathrel{=}\;\V{b}\mskip1.5mu\}\;\V{d})\;{}\<[28]%
\>[28]{}\mathrel{=}\;\anonymous \C{css}\;\{\mskip1.5mu \V{both}\;\mathrel{=}\;\V{b}\mskip1.5mu\}\;(\V{c}\;\F{+\!\!+}_C\;\V{d}){}\<[E]%
\\
\>[B]{}\V{c}\;\F{+\!\!+}_C\;\C{czz}\;{}\<[28]%
\>[28]{}\mathrel{=}\;\V{c}{}\<[E]%
\ColumnHook
\end{hscode}\resethooks
One way to build such a chunk is to observe that two scopes cover their concatenation.
\newpage 
\begin{hscode}\SaveRestoreHook
\column{B}{@{}>{\hspre}l<{\hspost}@{}}%
\column{19}{@{}>{\hspre}l<{\hspost}@{}}%
\column{29}{@{}>{\hspre}l<{\hspost}@{}}%
\column{37}{@{}>{\hspre}l<{\hspost}@{}}%
\column{58}{@{}>{\hspre}l<{\hspost}@{}}%
\column{62}{@{}>{\hspre}l<{\hspost}@{}}%
\column{91}{@{}>{\hspre}l<{\hspost}@{}}%
\column{E}{@{}>{\hspre}l<{\hspost}@{}}%
\>[B]{}\F{lrCop}\;\mathbin{:}\;{}\<[37]%
\>[37]{}(\V{iz}\;\V{jz}\;\mathbin{:}\;\D{Bwd}\;\V{K})\;\to \;{}\<[91]%
\>[91]{}\D{\Upsigma}\;(\V{iz}\;\D{\sqsubseteq}\;(\V{iz}\;\F{+\!\!+}\;\V{jz}))\;\lambda \;\V{\theta}\;\to \;\D{\Upsigma}\;(\V{jz}\;\D{\sqsubseteq}\;(\V{iz}\;\F{+\!\!+}\;\V{jz}))\;\lambda \;\V{\phi}\;\to \;\D{Cover}\;\V{ov}\;\V{\theta}\;\V{\phi}{}\<[E]%
\\
\>[B]{}\F{lrCop}\;\V{iz}\;{}\<[19]%
\>[19]{}(\V{jz}\;\C{\mbox{{}-}\!,}\;\V{j})\;{}\<[29]%
\>[29]{}\mathrel{=}\;\mathkw{let}\;\C{!}\;\C{!}\;\V{c}\;\mathrel{=}\;\F{lrCop}\;\V{iz}\;\V{jz}\;{}\<[58]%
\>[58]{}\mathkw{in}\;{}\<[62]%
\>[62]{}\C{!}\;\C{!}\;\V{c}\;\C{c\apo s}{}\<[E]%
\\
\>[B]{}\F{lrCop}\;(\V{iz}\;\C{\mbox{{}-}\!,}\;\V{i})\;{}\<[19]%
\>[19]{}\C{[]}\;{}\<[29]%
\>[29]{}\mathrel{=}\;\mathkw{let}\;\C{!}\;\C{!}\;\V{c}\;\mathrel{=}\;\F{lrCop}\;\V{iz}\;\C{[]}\;{}\<[58]%
\>[58]{}\mathkw{in}\;{}\<[62]%
\>[62]{}\C{!}\;\C{!}\;\V{c}\;\C{cs\apo}{}\<[E]%
\\
\>[B]{}\F{lrCop}\;\C{[]}\;{}\<[19]%
\>[19]{}\C{[]}\;{}\<[29]%
\>[29]{}\mathrel{=}\;{}\<[62]%
\>[62]{}\C{!}\;\C{!}\;\C{czz}{}\<[E]%
\ColumnHook
\end{hscode}\resethooks

Now, crucial to the enterprise is that the monoidal structure of scopes lets us not
only combine thinnings, but \emph{split} them, into global and local parts.
\begin{hscode}\SaveRestoreHook
\column{B}{@{}>{\hspre}l<{\hspost}@{}}%
\column{5}{@{}>{\hspre}l<{\hspost}@{}}%
\column{11}{@{}>{\hspre}l<{\hspost}@{}}%
\column{34}{@{}>{\hspre}l<{\hspost}@{}}%
\column{38}{@{}>{\hspre}l<{\hspost}@{}}%
\column{60}{@{}>{\hspre}l<{\hspost}@{}}%
\column{63}{@{}>{\hspre}l<{\hspost}@{}}%
\column{78}{@{}>{\hspre}l<{\hspost}@{}}%
\column{82}{@{}>{\hspre}l<{\hspost}@{}}%
\column{90}{@{}>{\hspre}l<{\hspost}@{}}%
\column{E}{@{}>{\hspre}l<{\hspost}@{}}%
\>[B]{}\anonymous \F{\dashv}\anonymous \;\mathbin{:}\;\forall\;{}\<[38]%
\>[38]{}\V{jz}\;{}\<[60]%
\>[60]{}(\V{\psi}\;\mathbin{:}\;\V{iz}\;\D{\sqsubseteq}\;(\V{kz}\;\F{+\!\!+}\;\V{jz}))\;\to \;{}\<[90]%
\>[90]{}\D{\Upsigma}\;\anonymous \;\lambda \;\V{kz'}\;\to \;\D{\Upsigma}\;\anonymous \;\lambda \;\V{jz'}\;\to {}\<[E]%
\\
\>[B]{}\hsindent{5}{}\<[5]%
\>[5]{}\D{\Upsigma}\;(\V{kz'}\;\D{\sqsubseteq}\;\V{kz})\;\lambda \;\V{\theta}\;\to \;\D{\Upsigma}\;(\V{jz'}\;\D{\sqsubseteq}\;\V{jz})\;\lambda \;\V{\phi}\;\to \;\D{\Upsigma}\;(\V{iz}\;\D{=\!\!\!\!=}\;(\V{kz'}\;\F{+\!\!+}\;\V{jz'}))\;\lambda \;{}\<[90]%
\>[90]{}\{\mskip1.5mu \C{refl}\;\to \;\V{\psi}\;\D{=\!\!\!\!=}\;(\V{\theta}\;\F{+\!\!+}_{\D{\sqsubseteq}}\;\V{\phi})\mskip1.5mu\}{}\<[E]%
\\[\blanklineskip]%
\>[B]{}\C{[]}\;{}\<[11]%
\>[11]{}\F{\dashv}\;\V{\psi}\;{}\<[63]%
\>[63]{}\mathrel{=}\;\C{!}\;\C{!}\;\V{\psi}\;\C{,}\;{}\<[78]%
\>[78]{}\C{oz}\;{}\<[82]%
\>[82]{}\C{,}\;\C{refl}\;\C{,}\;\C{refl}{}\<[E]%
\\
\>[B]{}(\V{jz}\;\C{\mbox{{}-}\!,}\;\V{j})\;{}\<[11]%
\>[11]{}\F{\dashv}\;(\V{\psi}\;\C{os})\;{}\<[34]%
\>[34]{}\mathkw{with}\;\V{jz}\;\F{\dashv}\;\V{\psi}{}\<[E]%
\\
\>[B]{}(\V{jz}\;\C{\mbox{{}-}\!,}\;\V{j})\;{}\<[11]%
\>[11]{}\F{\dashv}\;(.\;(\V{\theta}\;\F{+\!\!+}_{\D{\sqsubseteq}}\;\V{\phi})\;\C{os})\;{}\<[34]%
\>[34]{}\mid \;\C{!}\;\C{!}\;\V{\theta}\;\C{,}\;\V{\phi}\;\C{,}\;\C{refl}\;\C{,}\;\C{refl}\;{}\<[63]%
\>[63]{}\mathrel{=}\;\C{!}\;\C{!}\;\V{\theta}\;\C{,}\;\V{\phi}\;{}\<[78]%
\>[78]{}\C{os}\;{}\<[82]%
\>[82]{}\C{,}\;\C{refl}\;\C{,}\;\C{refl}{}\<[E]%
\\
\>[B]{}(\V{jz}\;\C{\mbox{{}-}\!,}\;\V{j})\;{}\<[11]%
\>[11]{}\F{\dashv}\;(\V{\psi}\;\C{o\apo})\;{}\<[34]%
\>[34]{}\mathkw{with}\;\V{jz}\;\F{\dashv}\;\V{\psi}{}\<[E]%
\\
\>[B]{}(\V{jz}\;\C{\mbox{{}-}\!,}\;\V{j})\;{}\<[11]%
\>[11]{}\F{\dashv}\;(.\;(\V{\theta}\;\F{+\!\!+}_{\D{\sqsubseteq}}\;\V{\phi})\;\C{o\apo})\;{}\<[34]%
\>[34]{}\mid \;\C{!}\;\C{!}\;\V{\theta}\;\C{,}\;\V{\phi}\;\C{,}\;\C{refl}\;\C{,}\;\C{refl}\;{}\<[63]%
\>[63]{}\mathrel{=}\;\C{!}\;\C{!}\;\V{\theta}\;\C{,}\;\V{\phi}\;{}\<[78]%
\>[78]{}\C{o\apo}\;{}\<[82]%
\>[82]{}\C{,}\;\C{refl}\;\C{,}\;\C{refl}{}\<[E]%
\ColumnHook
\end{hscode}\resethooks

Thus equipped, we can say how to bind some variables. The key is to say
at the binding site which of the bound variables will actually be used:
if they are not used, we should not even bring them into scope.

\noindent
\begin{tabular}{@{}l|l@{}}
\hspace*{ -0.3in}
\raisebox{0.7in}[0.6in][0in]{\parbox[t]{2.7in}{
\begin{hscode}\SaveRestoreHook
\column{B}{@{}>{\hspre}l<{\hspost}@{}}%
\column{3}{@{}>{\hspre}l<{\hspost}@{}}%
\column{25}{@{}>{\hspre}l<{\hspost}@{}}%
\column{35}{@{}>{\hspre}l<{\hspost}@{}}%
\column{45}{@{}>{\hspre}l<{\hspost}@{}}%
\column{E}{@{}>{\hspre}l<{\hspost}@{}}%
\>[B]{}\mathkw{data}\;\anonymous \D{\vdash}\anonymous \;{}\<[25]%
\>[25]{}\V{jz}\;(\V{T}\;\mathbin{:}\;\F{\overline{\black{\V{K}}}})\;\V{kz}\;\mathbin{:}\;\D{Set}\;\mathkw{where}{}\<[E]%
\\
\>[B]{}\hsindent{3}{}\<[3]%
\>[3]{}\anonymous \C{\fatbslash}\anonymous \;\mathbin{:}\;{}\<[35]%
\>[35]{}\V{iz}\;\D{\sqsubseteq}\;\V{jz}\;{}\<[45]%
\>[45]{}\to \;\V{T}\;(\V{kz}\;\F{+\!\!+}\;\V{iz})\;{}\<[E]%
\\
\>[45]{}\to \;(\V{jz}\;\D{\vdash}\;\V{T})\;\V{kz}{}\<[E]%
\ColumnHook
\end{hscode}\resethooks
}}
& \hspace*{ -0.4in}
\raisebox{0.7in}[0.6in][0in]{\parbox[t]{2.8in}{
\begin{hscode}\SaveRestoreHook
\column{B}{@{}>{\hspre}l<{\hspost}@{}}%
\column{29}{@{}>{\hspre}l<{\hspost}@{}}%
\column{37}{@{}>{\hspre}l<{\hspost}@{}}%
\column{59}{@{}>{\hspre}l<{\hspost}@{}}%
\column{E}{@{}>{\hspre}l<{\hspost}@{}}%
\>[B]{}\anonymous \F{\fatbslash}_R\anonymous \;\mathbin{:}\;\forall\;{}\<[37]%
\>[37]{}\V{jz}\;{}\<[59]%
\>[59]{}\to \;\V{T}\;\D{\Uparrow}\;(\V{kz}\;\F{+\!\!+}\;\V{jz})\;\to \;(\V{jz}\;\D{\vdash}\;\V{T})\;\D{\Uparrow}\;\V{kz}{}\<[E]%
\\
\>[B]{}\V{jz}\;\F{\fatbslash}_R\;(\V{t}\;\C{\uparrow}\;\V{\psi})\;{}\<[29]%
\>[29]{}\mathkw{with}\;\V{jz}\;\F{\dashv}\;\V{\psi}{}\<[E]%
\\
\>[B]{}\V{jz}\;\F{\fatbslash}_R\;(\V{t}\;\C{\uparrow}\;.\;(\V{\theta}\;\F{+\!\!+}_{\D{\sqsubseteq}}\;\V{\phi}))\;{}\<[29]%
\>[29]{}\mid \;\C{!}\;\C{!}\;\V{\theta}\;\C{,}\;\V{\phi}\;\C{,}\;\C{refl}\;\C{,}\;\C{refl}\;\mathrel{=}\;(\V{\phi}\;\C{\fatbslash}\;\V{t})\;\C{\uparrow}\;\V{\theta}{}\<[E]%
\ColumnHook
\end{hscode}\resethooks
}}
\end{tabular}

The monoid of scopes is generated from its singletons. By the time we \emph{use}
a variable, it should be the only thing in scope.
The associated smart constructor computes the thinned representation of variables.
\\
\parbox[t]{2.5in}{
\begin{hscode}\SaveRestoreHook
\column{B}{@{}>{\hspre}l<{\hspost}@{}}%
\column{3}{@{}>{\hspre}l<{\hspost}@{}}%
\column{23}{@{}>{\hspre}l<{\hspost}@{}}%
\column{E}{@{}>{\hspre}l<{\hspost}@{}}%
\>[B]{}\mathkw{data}\;\D{Va}_R\;{}\<[23]%
\>[23]{}(\V{k}\;\mathbin{:}\;\V{K})\;\mathbin{:}\;\F{\overline{\black{\V{K}}}}\;\mathkw{where}{}\<[E]%
\\
\>[B]{}\hsindent{3}{}\<[3]%
\>[3]{}\C{only}\;\mathbin{:}\;\D{Va}_R\;\V{k}\;(\C{[]}\;\C{\mbox{{}-}\!,}\;\V{k}){}\<[E]%
\ColumnHook
\end{hscode}\resethooks
}
\vrule
\parbox[t]{2.5in}{
\begin{hscode}\SaveRestoreHook
\column{B}{@{}>{\hspre}l<{\hspost}@{}}%
\column{46}{@{}>{\hspre}l<{\hspost}@{}}%
\column{E}{@{}>{\hspre}l<{\hspost}@{}}%
\>[B]{}\F{va}_R\;\mathbin{:}\;{}\<[46]%
\>[46]{}\V{k}\;\F{\leftarrow}\;\V{kz}\;\to \;\D{Va}_R\;\V{k}\;\D{\Uparrow}\;\V{kz}{}\<[E]%
\\
\>[B]{}\F{va}_R\;\V{x}\;\mathrel{=}\;\C{only}\;\C{\uparrow}\;\V{x}{}\<[E]%
\ColumnHook
\end{hscode}\resethooks
}

\vspace*{ -0.2in}
\paragraph{Untyped $\lambda$-calculus.~}
We can now give the $\lambda$-terms for which all \emph{free} variables are
relevant as follows.
Converting de Bruijn to co-de-Bruijn representation is easy with
smart constructors.
E.g., compare de Bruijn terms for the \ensuremath{\F{\mathbb{K}}} and \ensuremath{\F{\mathbb{S}}} combinators with their co-de-Bruijn form.
\vspace*{ -0.1in} \\
\parbox[t]{2.7in}{
\begin{hscode}\SaveRestoreHook
\column{B}{@{}>{\hspre}l<{\hspost}@{}}%
\column{3}{@{}>{\hspre}l<{\hspost}@{}}%
\column{8}{@{}>{\hspre}l<{\hspost}@{}}%
\column{31}{@{}>{\hspre}l<{\hspost}@{}}%
\column{36}{@{}>{\hspre}l<{\hspost}@{}}%
\column{E}{@{}>{\hspre}l<{\hspost}@{}}%
\>[B]{}\mathkw{data}\;\D{Lam}_R\;\mathbin{:}\;\F{\overline{\black{\D{One}}}}\;\mathkw{where}{}\<[E]%
\\
\>[B]{}\hsindent{3}{}\<[3]%
\>[3]{}\C{\scriptstyle\#}\;{}\<[8]%
\>[8]{}\mathbin{:}\;\D{Va}_R\;\C{\langle\rangle}\;{}\<[31]%
\>[31]{}\mathrel{\dot{\to}}\;{}\<[36]%
\>[36]{}\D{Lam}_R{}\<[E]%
\\
\>[B]{}\hsindent{3}{}\<[3]%
\>[3]{}\C{app}\;{}\<[8]%
\>[8]{}\mathbin{:}\;(\D{Lam}_R\;\D{\times}_R\;\D{Lam}_R)\;{}\<[31]%
\>[31]{}\mathrel{\dot{\to}}\;{}\<[36]%
\>[36]{}\D{Lam}_R{}\<[E]%
\\
\>[B]{}\hsindent{3}{}\<[3]%
\>[3]{}\C{\uplambda}\;{}\<[8]%
\>[8]{}\mathbin{:}\;(\C{[]}\;\C{\mbox{{}-}\!,}\;\C{\langle\rangle}\;\D{\vdash}\;\D{Lam}_R)\;{}\<[31]%
\>[31]{}\mathrel{\dot{\to}}\;{}\<[36]%
\>[36]{}\D{Lam}_R{}\<[E]%
\ColumnHook
\end{hscode}\resethooks
}
\vrule
\parbox[t]{3in}{
\begin{hscode}\SaveRestoreHook
\column{B}{@{}>{\hspre}l<{\hspost}@{}}%
\column{17}{@{}>{\hspre}l<{\hspost}@{}}%
\column{29}{@{}>{\hspre}l<{\hspost}@{}}%
\column{E}{@{}>{\hspre}l<{\hspost}@{}}%
\>[B]{}\F{lam}_R\;\mathbin{:}\;\D{Lam}\;\mathrel{\dot{\to}}\;(\D{Lam}_R\;\D{\Uparrow}\anonymous ){}\<[E]%
\\
\>[B]{}\F{lam}_R\;(\C{\scriptstyle\#}\;\V{x})\;{}\<[17]%
\>[17]{}\mathrel{=}\;\F{map\!\Uparrow}\;\C{\scriptstyle\#}\;{}\<[29]%
\>[29]{}(\F{va}_R\;\V{x}){}\<[E]%
\\
\>[B]{}\F{lam}_R\;(\V{f}\;\C{\scriptstyle\$}\;\V{s})\;{}\<[17]%
\>[17]{}\mathrel{=}\;\F{map\!\Uparrow}\;\C{app}\;{}\<[29]%
\>[29]{}(\F{lam}_R\;\V{f}\;\F{,}_R\;\F{lam}_R\;\V{s}){}\<[E]%
\\
\>[B]{}\F{lam}_R\;(\C{\uplambda}\;\V{t})\;{}\<[17]%
\>[17]{}\mathrel{=}\;\F{map\!\Uparrow}\;\C{\uplambda}\;{}\<[29]%
\>[29]{}(\anonymous \;\F{\fatbslash}_R\;\F{lam}_R\;\V{t}){}\<[E]%
\ColumnHook
\end{hscode}\resethooks
}
\vspace*{ -0.1in}
\begin{hscode}\SaveRestoreHook
\column{B}{@{}>{\hspre}l<{\hspost}@{}}%
\column{3}{@{}>{\hspre}l<{\hspost}@{}}%
\column{14}{@{}>{\hspre}l<{\hspost}@{}}%
\column{15}{@{}>{\hspre}l<{\hspost}@{}}%
\column{18}{@{}>{\hspre}l<{\hspost}@{}}%
\column{20}{@{}>{\hspre}l<{\hspost}@{}}%
\column{86}{@{}>{\hspre}l<{\hspost}@{}}%
\column{E}{@{}>{\hspre}l<{\hspost}@{}}%
\>[B]{}\F{\mathbb{K}}\;{}\<[15]%
\>[15]{}\mathrel{=}\;{}\<[18]%
\>[18]{}\C{\uplambda}\;(\C{\uplambda}\;(\C{\scriptstyle\#}\;(\F{oe}\;\C{os}\;\C{o\apo}))){}\<[E]%
\\
\>[B]{}\F{lam}_R\;\F{\mathbb{K}}\;{}\<[15]%
\>[15]{}\mathrel{=}\;{}\<[18]%
\>[18]{}\C{\uplambda}\;(\C{oz}\;\C{os}\;\C{\fatbslash}\;\C{\uplambda}\;(\C{oz}\;\C{o\apo}\;\C{\fatbslash}\;\C{\scriptstyle\#}\;\C{only}))\;\C{\uparrow}\;\C{oz}{}\<[E]%
\\[\blanklineskip]%
\>[B]{}\F{\mathbb{S}}\;{}\<[15]%
\>[15]{}\mathrel{=}\;{}\<[18]%
\>[18]{}\C{\uplambda}\;(\C{\uplambda}\;(\C{\uplambda}\;(\C{\scriptstyle\#}\;(\F{oe}\;\C{os}\;\C{o\apo}\;\C{o\apo})\;\C{\scriptstyle\$}\;\C{\scriptstyle\#}\;(\F{oe}\;\C{os})\;\C{\scriptstyle\$}\;(\C{\scriptstyle\#}\;(\F{oe}\;\C{os}\;\C{o\apo})\;\C{\scriptstyle\$}\;\C{\scriptstyle\#}\;(\F{oe}\;\C{os}))))){}\<[E]%
\\
\>[B]{}\F{lam}_R\;\F{\mathbb{S}}\;{}\<[15]%
\>[15]{}\mathrel{=}\;{}\<[18]%
\>[18]{}\C{\uplambda}\;(\C{oz}\;\C{os}\;\C{\fatbslash}\;\C{\uplambda}\;(\C{oz}\;\C{os}\;\C{\fatbslash}\;\C{\uplambda}\;(\C{oz}\;\C{os}\;\C{\fatbslash}{}\<[E]%
\\
\>[B]{}\hsindent{3}{}\<[3]%
\>[3]{}\C{app}\;(\C{pair}\;{}\<[14]%
\>[14]{}(\C{app}\;{}\<[20]%
\>[20]{}(\C{pair}\;(\C{\scriptstyle\#}\;\C{only}\;\C{\uparrow}\;\C{oz}\;\C{os}\;\C{o\apo})\;(\C{\scriptstyle\#}\;\C{only}\;\C{\uparrow}\;\C{oz}\;\C{o\apo}\;\C{os})\;(\C{czz}\;\C{cs\apo}\;\C{c\apo s}))\;{}\<[86]%
\>[86]{}\C{\uparrow}\;\C{oz}\;\C{os}\;\C{o\apo}\;\C{os})\;{}\<[E]%
\\
\>[14]{}(\C{app}\;{}\<[20]%
\>[20]{}(\C{pair}\;(\C{\scriptstyle\#}\;\C{only}\;\C{\uparrow}\;\C{oz}\;\C{os}\;\C{o\apo})\;(\C{\scriptstyle\#}\;\C{only}\;\C{\uparrow}\;\C{oz}\;\C{o\apo}\;\C{os})\;(\C{czz}\;\C{cs\apo}\;\C{c\apo s}))\;{}\<[86]%
\>[86]{}\C{\uparrow}\;\C{oz}\;\C{o\apo}\;\C{os}\;\C{os})\;{}\<[E]%
\\
\>[14]{}(\C{czz}\;\C{cs\apo}\;\C{c\apo s}\;\C{css})))))\;\C{\uparrow}\;\C{oz}{}\<[E]%
\ColumnHook
\end{hscode}\resethooks
Stare bravely! \ensuremath{\F{\mathbb{K}}} returns a plainly constant function.
Meanwhile,
\ensuremath{\F{\mathbb{S}}} clearly uses all three inputs: the function goes
left, the argument goes right, and the environment is shared.

\section{A Universe of Metasyntaxes-with-Binding}

There is nothing specific to the $\lambda$-calculus about de Bruijn
representation or its co-de-Bruijn counterpart. We may develop the
notions generically for multisorted syntaxes. If the sorts of our
syntax are drawn from set \ensuremath{\V{I}}, then we may characterize terms-with-binding
as inhabiting \ensuremath{\D{Kind}}s \ensuremath{\V{kz}\;\C{\Rightarrow}\;\V{i}}, which specify an extension of the scope
with new bindings \ensuremath{\V{kz}} and the sort \ensuremath{\V{i}} for the body of the binder.
\begin{hscode}\SaveRestoreHook
\column{B}{@{}>{\hspre}l<{\hspost}@{}}%
\column{36}{@{}>{\hspre}l<{\hspost}@{}}%
\column{E}{@{}>{\hspre}l<{\hspost}@{}}%
\>[B]{}\mathkw{record}\;\D{Kind}\;(\V{I}\;\mathbin{:}\;\D{Set})\;\mathbin{:}\;\D{Set}\;\mathkw{where}\;{}\<[36]%
\>[36]{}\mathkw{inductive};\quad\mathkw{constructor}\;\anonymous \C{\Rightarrow}\anonymous {}\<[E]%
\\
\>[36]{}\mathkw{field}\;\F{scope}\;\mathbin{:}\;\D{Bwd}\;(\D{Kind}\;\V{I});\quad\F{sort}\;\mathbin{:}\;\V{I}{}\<[E]%
\ColumnHook
\end{hscode}\resethooks
\ensuremath{\D{Kind}}s offer higher-order abstraction: a bound variable itself
has a \ensuremath{\D{Kind}}, being an object sort parametrized by a scope,
where the latter is, as in previous  sections, a \ensuremath{\D{Bwd}} list, with \ensuremath{\V{K}} now
fixed as \ensuremath{\D{Kind}\;\V{I}}. Object variables have sorts; \emph{meta}-variables have \ensuremath{\D{Kind}}s.
E.g., in the $\beta$-rule, $t$ and $s$ are not object variables like $x$
\[
  (\lambda x.\,t[x])\:s \;\leadsto\; t[s]
\]
but
placeholders, $s$ for some term and $t[x]$ for some term
with a parameter which can be and is instantiated, by $x$
on the left and $s$ on the right. The kind of $t$ is \ensuremath{\C{[]}\;\C{\mbox{{}-}\!,}\;(\C{[]}\;\C{\Rightarrow}\;\C{\langle\rangle})\;\C{\Rightarrow}\;\C{\langle\rangle}}.

We may give the syntax of each sort as a function mapping sorts to
\ensuremath{\D{Desc}}riptions
\ensuremath{\V{D}\;\mathbin{:}\;\V{I}\;\to \;\D{Desc}\;\V{I}}.
\begin{hscode}\SaveRestoreHook
\column{B}{@{}>{\hspre}l<{\hspost}@{}}%
\column{3}{@{}>{\hspre}l<{\hspost}@{}}%
\column{10}{@{}>{\hspre}l<{\hspost}@{}}%
\column{23}{@{}>{\hspre}l<{\hspost}@{}}%
\column{32}{@{}>{\hspre}l<{\hspost}@{}}%
\column{39}{@{}>{\hspre}l<{\hspost}@{}}%
\column{80}{@{}>{\hspre}l<{\hspost}@{}}%
\column{E}{@{}>{\hspre}l<{\hspost}@{}}%
\>[B]{}\mathkw{data}\;\D{Desc}\;(\V{I}\;\mathbin{:}\;\D{Set})\;\mathbin{:}\;\D{Set}_1\;\mathkw{where}{}\<[E]%
\\
\>[B]{}\hsindent{3}{}\<[3]%
\>[3]{}\C{Rec}_D\;{}\<[10]%
\>[10]{}\mathbin{:}\;\D{Kind}\;\V{I}\;\to \;{}\<[23]%
\>[23]{}\D{Desc}\;\V{I};\quad{}\<[32]%
\>[32]{}\C{\Upsigma}_D\;{}\<[39]%
\>[39]{}\mathbin{:}\;(\V{S}\;\mathbin{:}\;\D{Datoid})\;\to \;(\F{Data}\;\V{S}\;\to \;\D{Desc}\;\V{I})\;\to \;{}\<[80]%
\>[80]{}\D{Desc}\;\V{I}{}\<[E]%
\\
\>[B]{}\hsindent{3}{}\<[3]%
\>[3]{}\C{One}_D\;{}\<[10]%
\>[10]{}\mathbin{:}\;{}\<[23]%
\>[23]{}\D{Desc}\;\V{I};\quad{}\<[32]%
\>[32]{}\anonymous \C{\times}_D\anonymous \;{}\<[39]%
\>[39]{}\mathbin{:}\;\D{Desc}\;\V{I}\;\to \;\D{Desc}\;\V{I}\;\to \;{}\<[80]%
\>[80]{}\D{Desc}\;\V{I}{}\<[E]%
\ColumnHook
\end{hscode}\resethooks
We may ask for a subterm with a given \ensuremath{\D{Kind}}, so it can bind
variables by listing their \ensuremath{\D{Kind}}s left of \ensuremath{\C{\Rightarrow}}. Descriptions
are closed under unit and pairing.
We may also ask
for terms to be tagged by some sort of `constructor' inhabiting
some \ensuremath{\D{Datoid}}, i.e., a set with a decidable equality, given
as follows:
\vspace*{ -0.1in}\\
\parbox[t]{2.7in}{
\begin{hscode}\SaveRestoreHook
\column{B}{@{}>{\hspre}l<{\hspost}@{}}%
\column{3}{@{}>{\hspre}l<{\hspost}@{}}%
\column{8}{@{}>{\hspre}l<{\hspost}@{}}%
\column{26}{@{}>{\hspre}l<{\hspost}@{}}%
\column{E}{@{}>{\hspre}l<{\hspost}@{}}%
\>[B]{}\mathkw{data}\;\D{Decide}\;(\V{X}\;\mathbin{:}\;\D{Set})\;\mathbin{:}\;\D{Set}\;\mathkw{where}{}\<[E]%
\\
\>[B]{}\hsindent{3}{}\<[3]%
\>[3]{}\C{yes}\;{}\<[8]%
\>[8]{}\mathbin{:}\;\V{X}\;\to \;{}\<[26]%
\>[26]{}\D{Decide}\;\V{X}{}\<[E]%
\\
\>[B]{}\hsindent{3}{}\<[3]%
\>[3]{}\C{no}\;{}\<[8]%
\>[8]{}\mathbin{:}\;(\V{X}\;\to \;\D{Zero})\;\to \;{}\<[26]%
\>[26]{}\D{Decide}\;\V{X}{}\<[E]%
\ColumnHook
\end{hscode}\resethooks
}
\vrule
\parbox[t]{3in}{
\begin{hscode}\SaveRestoreHook
\column{B}{@{}>{\hspre}l<{\hspost}@{}}%
\column{3}{@{}>{\hspre}l<{\hspost}@{}}%
\column{10}{@{}>{\hspre}l<{\hspost}@{}}%
\column{18}{@{}>{\hspre}l<{\hspost}@{}}%
\column{E}{@{}>{\hspre}l<{\hspost}@{}}%
\>[B]{}\mathkw{record}\;\D{Datoid}\;\mathbin{:}\;\D{Set}_1\;\mathkw{where}{}\<[E]%
\\
\>[B]{}\hsindent{3}{}\<[3]%
\>[3]{}\mathkw{field}\;{}\<[10]%
\>[10]{}\F{Data}\;{}\<[18]%
\>[18]{}\mathbin{:}\;\D{Set}{}\<[E]%
\\
\>[10]{}\F{decide}\;{}\<[18]%
\>[18]{}\mathbin{:}\;(\V{x}\;\V{y}\;\mathbin{:}\;\F{Data})\;\to \;\D{Decide}\;(\V{x}\;\D{=\!\!\!\!=}\;\V{y}){}\<[E]%
\ColumnHook
\end{hscode}\resethooks
}

\paragraph{Describing untyped $\lambda$-calculus.~} Define a tag enumeration, then a description.
\vspace*{ -0.1in} \\
\parbox[t]{3.2in}{
\begin{hscode}\SaveRestoreHook
\column{B}{@{}>{\hspre}l<{\hspost}@{}}%
\column{9}{@{}>{\hspre}l<{\hspost}@{}}%
\column{E}{@{}>{\hspre}l<{\hspost}@{}}%
\>[B]{}\mathkw{data}\;\D{LamTag}\;\mathbin{:}\;\D{Set}\;\mathkw{where}\;\C{app}\;\C{\uplambda}\;\mathbin{:}\;\D{LamTag}{}\<[E]%
\\[\blanklineskip]%
\>[B]{}\F{LAMTAG}\;\mathbin{:}\;\D{Datoid}{}\<[E]%
\\
\>[B]{}\F{Data}\;{}\<[9]%
\>[9]{}\F{LAMTAG}\;\mathrel{=}\;\D{LamTag}{}\<[E]%
\ColumnHook
\end{hscode}\resethooks
}
\vrule
\hspace*{ -0.3in}
\parbox[t]{2.5in}{
\begin{hscode}\SaveRestoreHook
\column{B}{@{}>{\hspre}l<{\hspost}@{}}%
\column{9}{@{}>{\hspre}l<{\hspost}@{}}%
\column{21}{@{}>{\hspre}l<{\hspost}@{}}%
\column{26}{@{}>{\hspre}l<{\hspost}@{}}%
\column{E}{@{}>{\hspre}l<{\hspost}@{}}%
\>[B]{}\F{decide}\;{}\<[9]%
\>[9]{}\F{LAMTAG}\;\C{app}\;{}\<[21]%
\>[21]{}\C{app}\;{}\<[26]%
\>[26]{}\mathrel{=}\;\C{yes}\;\C{refl}{}\<[E]%
\\
\>[B]{}\F{decide}\;{}\<[9]%
\>[9]{}\F{LAMTAG}\;\C{app}\;{}\<[21]%
\>[21]{}\C{\uplambda}\;{}\<[26]%
\>[26]{}\mathrel{=}\;\C{no}\;\lambda \;(){}\<[E]%
\\
\>[B]{}\F{decide}\;{}\<[9]%
\>[9]{}\F{LAMTAG}\;\C{\uplambda}\;{}\<[21]%
\>[21]{}\C{app}\;{}\<[26]%
\>[26]{}\mathrel{=}\;\C{no}\;\lambda \;(){}\<[E]%
\\
\>[B]{}\F{decide}\;{}\<[9]%
\>[9]{}\F{LAMTAG}\;\C{\uplambda}\;{}\<[21]%
\>[21]{}\C{\uplambda}\;{}\<[26]%
\>[26]{}\mathrel{=}\;\C{yes}\;\C{refl}{}\<[E]%
\ColumnHook
\end{hscode}\resethooks
}
\vspace*{ -0.3in}
\begin{hscode}\SaveRestoreHook
\column{B}{@{}>{\hspre}l<{\hspost}@{}}%
\column{25}{@{}>{\hspre}l<{\hspost}@{}}%
\column{32}{@{}>{\hspre}l<{\hspost}@{}}%
\column{E}{@{}>{\hspre}l<{\hspost}@{}}%
\>[B]{}\F{Lam}_D\;\mathbin{:}\;\D{One}\;\to \;\D{Desc}\;\D{One}{}\<[E]%
\\
\>[B]{}\F{Lam}_D\;\C{\langle\rangle}\;\mathrel{=}\;\C{\Upsigma}_D\;\F{LAMTAG}\;\lambda \;{}\<[25]%
\>[25]{}\{\mskip1.5mu \C{app}\;{}\<[32]%
\>[32]{}\to \;\C{Rec}_D\;(\C{[]}\;\C{\Rightarrow}\;\C{\langle\rangle})\;\C{\times}_D\;\C{Rec}_D\;(\C{[]}\;\C{\Rightarrow}\;\C{\langle\rangle}){}\<[E]%
\\
\>[25]{};\quad\C{\uplambda}\;{}\<[32]%
\>[32]{}\to \;\C{Rec}_D\;(\C{[]}\;\C{\mbox{{}-}\!,}\;(\C{[]}\;\C{\Rightarrow}\;\C{\langle\rangle})\;\C{\Rightarrow}\;\C{\langle\rangle})\mskip1.5mu\}{}\<[E]%
\ColumnHook
\end{hscode}\resethooks
Note that we do not and cannot include a tag or description for
the use sites of variables in terms: use of variables in scope
pertains not to the specific syntax, but to the general notion
of what it is to be a syntax.

\paragraph{Interpreting \ensuremath{\D{Desc}} as de Bruijn Syntax.~}
Let us give the de Bruijn interpretation of our syntax descriptions.
We give meaning to \ensuremath{\D{Desc}} in the traditional manner, interpreting
them as strictly positive operators in some \ensuremath{\V{R}} which gives the semantics
to \ensuremath{\C{Rec}_D}. In recursive positions, the scope grows by the bindings demanded by the given \ensuremath{\D{Kind}}.
At use sites, higher-kinded variables must be instantiated, just
like $t[x]$ in the $\beta$-rule example: \ensuremath{\F{\overrightarrow{\black{\cdot }}}} computes
the \ensuremath{\D{Desc}}ription of the
spine of actual parameters required.
\vspace*{ -0.1in} \\
\parbox[t]{3.8in}{
\begin{hscode}\SaveRestoreHook
\column{B}{@{}>{\hspre}l<{\hspost}@{}}%
\column{13}{@{}>{\hspre}l<{\hspost}@{}}%
\column{E}{@{}>{\hspre}l<{\hspost}@{}}%
\>[B]{}\F{\llbracket}\anonymous \F{\mid}\anonymous \F{\rrbracket}\;\mathbin{:}\;\forall\;\{\mskip1.5mu \V{I}\mskip1.5mu\}\;\to \;\D{Desc}\;\V{I}\;\to \;(\V{I}\;\to \;\F{\overline{\black{\D{Kind}\;\V{I}}}})\;\to \;\F{\overline{\black{\D{Kind}\;\V{I}}}}{}\<[E]%
\\
\>[B]{}\F{\llbracket}\;\C{Rec}_D\;\V{k}\;{}\<[13]%
\>[13]{}\F{\mid}\;\V{R}\;\F{\rrbracket}\;\V{kz}\;\mathrel{=}\;\V{R}\;(\F{sort}\;\V{k})\;(\V{kz}\;\F{+\!\!+}\;\F{scope}\;\V{k}){}\<[E]%
\\
\>[B]{}\F{\llbracket}\;\C{\Upsigma}_D\;\V{S}\;\V{T}\;{}\<[13]%
\>[13]{}\F{\mid}\;\V{R}\;\F{\rrbracket}\;\V{kz}\;\mathrel{=}\;\D{\Upsigma}\;(\F{Data}\;\V{S})\;\lambda \;\V{s}\;\to \;\F{\llbracket}\;\V{T}\;\V{s}\;\F{\mid}\;\V{R}\;\F{\rrbracket}\;\V{kz}{}\<[E]%
\\
\>[B]{}\F{\llbracket}\;\C{One}_D\;{}\<[13]%
\>[13]{}\F{\mid}\;\V{R}\;\F{\rrbracket}\;\V{kz}\;\mathrel{=}\;\D{One}{}\<[E]%
\\
\>[B]{}\F{\llbracket}\;\V{S}\;\C{\times}_D\;\V{T}\;{}\<[13]%
\>[13]{}\F{\mid}\;\V{R}\;\F{\rrbracket}\;\V{kz}\;\mathrel{=}\;\F{\llbracket}\;\V{S}\;\F{\mid}\;\V{R}\;\F{\rrbracket}\;\V{kz}\;\F{\times}\;\F{\llbracket}\;\V{T}\;\F{\mid}\;\V{R}\;\F{\rrbracket}\;\V{kz}{}\<[E]%
\ColumnHook
\end{hscode}\resethooks
}
\vrule
\hspace*{ -0.3in}
\parbox[t]{2in}{
\begin{hscode}\SaveRestoreHook
\column{B}{@{}>{\hspre}l<{\hspost}@{}}%
\column{15}{@{}>{\hspre}l<{\hspost}@{}}%
\column{30}{@{}>{\hspre}l<{\hspost}@{}}%
\column{32}{@{}>{\hspre}l<{\hspost}@{}}%
\column{36}{@{}>{\hspre}l<{\hspost}@{}}%
\column{E}{@{}>{\hspre}l<{\hspost}@{}}%
\>[B]{}\F{\overrightarrow{\black{\raisebox{0in}[0.07in][0in]{\_}}}}\;\mathbin{:}\;{}\<[32]%
\>[32]{}\D{Bwd}\;(\D{Kind}\;\V{I})\;\to \;\D{Desc}\;\V{I}{}\<[E]%
\\
\>[B]{}\F{\overrightarrow{\black{\C{[]}}}}\;{}\<[15]%
\>[15]{}\mathrel{=}\;\C{One}_D{}\<[E]%
\\
\>[B]{}\F{\overrightarrow{\black{\V{kz}\;\C{\mbox{{}-}\!,}\;\V{k}}}}\;{}\<[15]%
\>[15]{}\mathrel{=}\;\F{\overrightarrow{\black{\V{kz}}}}\;{}\<[30]%
\>[30]{}\C{\times}_D\;{}\<[36]%
\>[36]{}\C{Rec}_D\;\V{k}{}\<[E]%
\ColumnHook
\end{hscode}\resethooks
}

Tying the knot, we find that a term is either a variable instantiated
with its spine of actual parameters, or it is a construct of the syntax
for the demanded sort, with subterms in recursive positions.
\begin{hscode}\SaveRestoreHook
\column{B}{@{}>{\hspre}l<{\hspost}@{}}%
\column{3}{@{}>{\hspre}l<{\hspost}@{}}%
\column{9}{@{}>{\hspre}l<{\hspost}@{}}%
\column{22}{@{}>{\hspre}l<{\hspost}@{}}%
\column{36}{@{}>{\hspre}l<{\hspost}@{}}%
\column{56}{@{}>{\hspre}l<{\hspost}@{}}%
\column{69}{@{}>{\hspre}l<{\hspost}@{}}%
\column{87}{@{}>{\hspre}l<{\hspost}@{}}%
\column{E}{@{}>{\hspre}l<{\hspost}@{}}%
\>[B]{}\mathkw{data}\;\D{Tm}\;{}\<[22]%
\>[22]{}(\V{D}\;\mathbin{:}\;\V{I}\;\to \;\D{Desc}\;\V{I})\;(\V{i}\;\mathbin{:}\;\V{I})\;\V{kz}\;\mathbin{:}\;\D{Set}\;\mathkw{where}\;\mbox{\onelinecomment  \ensuremath{\D{Tm}\;\V{D}\;\V{i}\;\mathbin{:}\;\F{\overline{\black{\D{Kind}\;\V{I}}}}}}{}\<[E]%
\\
\>[B]{}\hsindent{3}{}\<[3]%
\>[3]{}\anonymous \C{\scriptstyle \#\$}\anonymous \;{}\<[9]%
\>[9]{}\mathbin{:}\;{}\<[36]%
\>[36]{}(\V{jz}\;\C{\Rightarrow}\;\V{i})\;\F{\leftarrow}\;\V{kz}\;\to \;{}\<[56]%
\>[56]{}\F{\llbracket}\;\F{\overrightarrow{\black{\V{jz}}}}\;{}\<[69]%
\>[69]{}\F{\mid}\;\D{Tm}\;\V{D}\;\F{\rrbracket}\;\V{kz}\;\to \;{}\<[87]%
\>[87]{}\D{Tm}\;\V{D}\;\V{i}\;\V{kz}{}\<[E]%
\\
\>[B]{}\hsindent{3}{}\<[3]%
\>[3]{}\C{[}\anonymous \C{]}\;{}\<[9]%
\>[9]{}\mathbin{:}\;{}\<[56]%
\>[56]{}\F{\llbracket}\;\V{D}\;\V{i}\;{}\<[69]%
\>[69]{}\F{\mid}\;\D{Tm}\;\V{D}\;\F{\rrbracket}\;\V{kz}\;\to \;{}\<[87]%
\>[87]{}\D{Tm}\;\V{D}\;\V{i}\;\V{kz}{}\<[E]%
\ColumnHook
\end{hscode}\resethooks

\vspace*{ -0.2in}
\paragraph{Interpreting \ensuremath{\D{Desc}} as co-de-Bruijn Syntax.~}
Now let us interpret \ensuremath{\D{Desc}}riptions in co-de-Bruijn style,
enforcing that all variables in scope are relevant, and that
binding sites expose vacuity.
\begin{hscode}\SaveRestoreHook
\column{B}{@{}>{\hspre}l<{\hspost}@{}}%
\column{3}{@{}>{\hspre}l<{\hspost}@{}}%
\column{9}{@{}>{\hspre}l<{\hspost}@{}}%
\column{13}{@{}>{\hspre}l<{\hspost}@{}}%
\column{23}{@{}>{\hspre}l<{\hspost}@{}}%
\column{36}{@{}>{\hspre}l<{\hspost}@{}}%
\column{37}{@{}>{\hspre}l<{\hspost}@{}}%
\column{55}{@{}>{\hspre}l<{\hspost}@{}}%
\column{68}{@{}>{\hspre}l<{\hspost}@{}}%
\column{83}{@{}>{\hspre}l<{\hspost}@{}}%
\column{88}{@{}>{\hspre}l<{\hspost}@{}}%
\column{E}{@{}>{\hspre}l<{\hspost}@{}}%
\>[B]{}\F{\llbracket}\anonymous \F{\mid}\anonymous \F{\rrbracket}_R\;\mathbin{:}\;{}\<[37]%
\>[37]{}\D{Desc}\;\V{I}\;\to \;(\V{I}\;\to \;\F{\overline{\black{\D{Kind}\;\V{I}}}})\;\to \;\F{\overline{\black{\D{Kind}\;\V{I}}}}{}\<[E]%
\\
\>[B]{}\F{\llbracket}\;\C{Rec}_D\;\V{k}\;{}\<[13]%
\>[13]{}\F{\mid}\;\V{R}\;\F{\rrbracket}_R\;\mathrel{=}\;\F{scope}\;\V{k}\;\D{\vdash}\;\V{R}\;(\F{sort}\;\V{k}){}\<[E]%
\\
\>[B]{}\F{\llbracket}\;\C{\Upsigma}_D\;\V{S}\;\V{T}\;{}\<[13]%
\>[13]{}\F{\mid}\;\V{R}\;\F{\rrbracket}_R\;\mathrel{=}\;\lambda \;\V{kz}\;\to \;\D{\Upsigma}\;(\F{Data}\;\V{S})\;\lambda \;\V{s}\;\to \;\F{\llbracket}\;\V{T}\;\V{s}\;\F{\mid}\;\V{R}\;\F{\rrbracket}_R\;\V{kz}{}\<[E]%
\\
\>[B]{}\F{\llbracket}\;\C{One}_D\;{}\<[13]%
\>[13]{}\F{\mid}\;\V{R}\;\F{\rrbracket}_R\;\mathrel{=}\;\D{One}_{R}{}\<[E]%
\\
\>[B]{}\F{\llbracket}\;\V{S}\;\C{\times}_D\;\V{T}\;{}\<[13]%
\>[13]{}\F{\mid}\;\V{R}\;\F{\rrbracket}_R\;\mathrel{=}\;\F{\llbracket}\;\V{S}\;\F{\mid}\;\V{R}\;\F{\rrbracket}_R\;\D{\times}_R\;\F{\llbracket}\;\V{T}\;\F{\mid}\;\V{R}\;\F{\rrbracket}_R{}\<[E]%
\\[\blanklineskip]%
\>[B]{}\mathkw{data}\;\D{Tm}_R\;{}\<[23]%
\>[23]{}(\V{D}\;\mathbin{:}\;\V{I}\;\to \;\D{Desc}\;\V{I})\;(\V{i}\;\mathbin{:}\;\V{I})\;\mathbin{:}\;\F{\overline{\black{\D{Kind}\;\V{I}}}}\;\mathkw{where}{}\<[E]%
\\
\>[B]{}\hsindent{3}{}\<[3]%
\>[3]{}\C{\scriptstyle \#}\;{}\<[9]%
\>[9]{}\mathbin{:}\;{}\<[36]%
\>[36]{}(\D{Va}_R\;(\V{jz}\;\C{\Rightarrow}\;\V{i})\;\D{\times}_R\;{}\<[55]%
\>[55]{}\F{\llbracket}\;\F{\overrightarrow{\black{\V{jz}}}}\;{}\<[68]%
\>[68]{}\F{\mid}\;\D{Tm}_R\;\V{D}\;\F{\rrbracket}_R)\;{}\<[83]%
\>[83]{}\mathrel{\dot{\to}}\;{}\<[88]%
\>[88]{}\D{Tm}_R\;\V{D}\;\V{i}{}\<[E]%
\\
\>[B]{}\hsindent{3}{}\<[3]%
\>[3]{}\C{[}\anonymous \C{]}\;{}\<[9]%
\>[9]{}\mathbin{:}\;{}\<[55]%
\>[55]{}\F{\llbracket}\;\V{D}\;\V{i}\;{}\<[68]%
\>[68]{}\F{\mid}\;\D{Tm}_R\;\V{D}\;\F{\rrbracket}_R\;{}\<[83]%
\>[83]{}\mathrel{\dot{\to}}\;{}\<[88]%
\>[88]{}\D{Tm}_R\;\V{D}\;\V{i}{}\<[E]%
\ColumnHook
\end{hscode}\resethooks

We can compute co-de-Bruijn terms from de Bruijn terms, generically.
\begin{hscode}\SaveRestoreHook
\column{B}{@{}>{\hspre}l<{\hspost}@{}}%
\column{8}{@{}>{\hspre}l<{\hspost}@{}}%
\column{25}{@{}>{\hspre}l<{\hspost}@{}}%
\column{43}{@{}>{\hspre}l<{\hspost}@{}}%
\column{48}{@{}>{\hspre}l<{\hspost}@{}}%
\column{52}{@{}>{\hspre}l<{\hspost}@{}}%
\column{55}{@{}>{\hspre}l<{\hspost}@{}}%
\column{56}{@{}>{\hspre}l<{\hspost}@{}}%
\column{73}{@{}>{\hspre}l<{\hspost}@{}}%
\column{78}{@{}>{\hspre}l<{\hspost}@{}}%
\column{E}{@{}>{\hspre}l<{\hspost}@{}}%
\>[B]{}\F{code}\;{}\<[8]%
\>[8]{}\mathbin{:}\;{}\<[56]%
\>[56]{}\D{Tm}\;\V{D}\;\V{i}\;{}\<[73]%
\>[73]{}\mathrel{\dot{\to}}\;{}\<[78]%
\>[78]{}(\D{Tm}_R\;\V{D}\;\V{i}\;\D{\Uparrow}\anonymous ){}\<[E]%
\\
\>[B]{}\F{codes}\;{}\<[8]%
\>[8]{}\mathbin{:}\;{}\<[48]%
\>[48]{}\V{S}\;{}\<[52]%
\>[52]{}\to \;{}\<[56]%
\>[56]{}\F{\llbracket}\;\V{S}\;\F{\mid}\;\D{Tm}\;\V{D}\;\F{\rrbracket}\;{}\<[73]%
\>[73]{}\mathrel{\dot{\to}}\;{}\<[78]%
\>[78]{}(\F{\llbracket}\;\V{S}\;\F{\mid}\;\D{Tm}_R\;\V{D}\;\F{\rrbracket}_R\;\D{\Uparrow}\anonymous ){}\<[E]%
\\
\>[B]{}\F{code}\;{}\<[25]%
\>[25]{}(\anonymous \C{\scriptstyle \#\$}\anonymous \;\{\mskip1.5mu \V{jz}\mskip1.5mu\}\;\V{x}\;\V{ts})\;{}\<[43]%
\>[43]{}\mathrel{=}\;\F{map\!\Uparrow}\;\C{\scriptstyle \#}\;{}\<[55]%
\>[55]{}(\F{va}_R\;\V{x}\;\F{,}_R\;\F{codes}\;\F{\overrightarrow{\black{\V{jz}}}}\;\V{ts}){}\<[E]%
\\
\>[B]{}\F{code}\;\{\mskip1.5mu \V{D}\;\mathrel{=}\;\V{D}\mskip1.5mu\}\;\{\mskip1.5mu \V{i}\;\mathrel{=}\;\V{i}\mskip1.5mu\}\;{}\<[25]%
\>[25]{}\C{[}\;\V{ts}\;\C{]}\;{}\<[43]%
\>[43]{}\mathrel{=}\;\F{map\!\Uparrow}\;\C{[}\anonymous \C{]}\;{}\<[55]%
\>[55]{}(\F{codes}\;(\V{D}\;\V{i})\;\V{ts}){}\<[E]%
\\
\>[B]{}\F{codes}\;(\C{Rec}_D\;\V{k})\;{}\<[25]%
\>[25]{}\V{t}\;{}\<[43]%
\>[43]{}\mathrel{=}\;\F{scope}\;\V{k}\;\F{\fatbslash}_R\;\F{code}\;\V{t}{}\<[E]%
\\
\>[B]{}\F{codes}\;(\C{\Upsigma}_D\;\V{S}\;\V{T})\;{}\<[25]%
\>[25]{}(\V{s}\;\C{,}\;\V{ts})\;{}\<[43]%
\>[43]{}\mathrel{=}\;\F{map\!\Uparrow}\;(\V{s}\;\V{,\char95 })\;(\F{codes}\;(\V{T}\;\V{s})\;\V{ts}){}\<[E]%
\\
\>[B]{}\F{codes}\;\C{One}_D\;{}\<[25]%
\>[25]{}\C{\langle\rangle}\;{}\<[43]%
\>[43]{}\mathrel{=}\;\F{\langle\rangle}_{R}{}\<[E]%
\\
\>[B]{}\F{codes}\;(\V{S}\;\C{\times}_D\;\V{T})\;{}\<[25]%
\>[25]{}(\V{ss}\;\C{,}\;\V{ts})\;{}\<[43]%
\>[43]{}\mathrel{=}\;\F{codes}\;\V{S}\;\V{ss}\;\F{,}_R\;\F{codes}\;\V{T}\;\V{ts}{}\<[E]%
\ColumnHook
\end{hscode}\resethooks
The reverse translation is left as an (easy) exercise in thinning composition for the reader.

\section{Hereditary Substitution for Co-de-Bruijn Metasyntax}

Let us develop the appropriate notion of substitution for our metasyntax, \emph{hereditary}
in the sense of Watkins et al.~\cite{DBLP:conf/types/WatkinsCPW03}. Substituting a higher-kinded
variable requires us further to substitute its parameters.

We shall need a type to represent the fate of each variable in some source scope as we
construct a term in some target scope. I call this type \ensuremath{\D{HSub}}:
let us work through it slowly.
\begin{hscode}\SaveRestoreHook
\column{B}{@{}>{\hspre}l<{\hspost}@{}}%
\column{3}{@{}>{\hspre}l<{\hspost}@{}}%
\column{18}{@{}>{\hspre}l<{\hspost}@{}}%
\column{28}{@{}>{\hspre}l<{\hspost}@{}}%
\column{45}{@{}>{\hspre}l<{\hspost}@{}}%
\column{E}{@{}>{\hspre}l<{\hspost}@{}}%
\>[B]{}\mathkw{record}\;\D{HSub}\;\{\mskip1.5mu \V{I}\mskip1.5mu\}\;{}\<[18]%
\>[18]{}(\V{D}\;{}\<[28]%
\>[28]{}\mathbin{:}\;\V{I}\;\to \;\D{Desc}\;\V{I})\;{}\<[45]%
\>[45]{}\mbox{\onelinecomment  the underlying syntax}{}\<[E]%
\\
\>[18]{}(\V{src}\;\V{trg}\;{}\<[28]%
\>[28]{}\mathbin{:}\;\D{Bwd}\;(\D{Kind}\;\V{I}))\;{}\<[45]%
\>[45]{}\mbox{\onelinecomment  source and target scopes}{}\<[E]%
\\
\>[18]{}(\V{act}\;{}\<[28]%
\>[28]{}\mathbin{:}\;\D{Bwd}\;(\D{Kind}\;\V{I}))\;{}\<[45]%
\>[45]{}\mbox{\onelinecomment  the \emph{active} subscope}{}\<[E]%
\\
\>[18]{}\mathbin{:}\;\D{Set}\;\mathkw{where}{}\<[E]%
\\
\>[B]{}\hsindent{3}{}\<[3]%
\>[3]{}\mathkw{constructor}\;\anonymous \C{\sqsubseteq\![}\anonymous \C{]\!:\!=}\anonymous {}\<[E]%
\\
\>[B]{}\hsindent{3}{}\<[3]%
\>[3]{}\mathkw{field}\;\mbox{\onelinecomment  to follow}{}\<[E]%
\ColumnHook
\end{hscode}\resethooks
While \ensuremath{\V{D}}, \ensuremath{\V{src}} and \ensuremath{\V{trg}} indicate the task at hand, the extra scope parameter,
\ensuremath{\V{act}}, serves a more subtle purpose: let us see how, presently.
The mixfix constructor is intended to suggest that the \ensuremath{\F{parti}}tion in the middle
splits the source scope into \ensuremath{\F{passive}} and \ensuremath{\F{active}} variables, with different fates,
respectively, thinning into the target scope and actual substitution:
\begin{hscode}\SaveRestoreHook
\column{B}{@{}>{\hspre}l<{\hspost}@{}}%
\column{5}{@{}>{\hspre}l<{\hspost}@{}}%
\column{16}{@{}>{\hspre}l<{\hspost}@{}}%
\column{24}{@{}>{\hspre}l<{\hspost}@{}}%
\column{50}{@{}>{\hspre}l<{\hspost}@{}}%
\column{E}{@{}>{\hspre}l<{\hspost}@{}}%
\>[5]{}\{\mskip1.5mu \F{pass}\mskip1.5mu\}\;{}\<[16]%
\>[16]{}\mathbin{:}\;\D{Bwd}\;(\D{Kind}\;\V{I}){}\<[E]%
\\
\>[5]{}\{\mskip1.5mu \F{passive}\mskip1.5mu\}\;{}\<[16]%
\>[16]{}\mathbin{:}\;\F{pass}\;{}\<[24]%
\>[24]{}\D{\sqsubseteq}\;\V{src}{}\<[E]%
\\
\>[5]{}\{\mskip1.5mu \F{active}\mskip1.5mu\}\;{}\<[16]%
\>[16]{}\mathbin{:}\;\V{act}\;{}\<[24]%
\>[24]{}\D{\sqsubseteq}\;\V{src}{}\<[E]%
\\
\>[5]{}\F{passTrg}\;{}\<[16]%
\>[16]{}\mathbin{:}\;\F{pass}\;{}\<[24]%
\>[24]{}\D{\sqsubseteq}\;\V{trg}{}\<[50]%
\>[50]{}\mbox{\onelinecomment  \ensuremath{\F{passive}} variables are `renamed'}{}\<[E]%
\\
\>[5]{}\F{parti}\;{}\<[16]%
\>[16]{}\mathbin{:}\;\D{Cover}\;\C{f\!f}\;\F{passive}\;\F{active}{}\<[50]%
\>[50]{}\mbox{\onelinecomment  \ensuremath{\C{f\!f}} forbids overlap}{}\<[E]%
\\
\>[5]{}\F{images}\;{}\<[16]%
\>[16]{}\mathbin{:}\;\F{\llbracket}\;\F{\overrightarrow{\black{\cdot }}}\;\V{act}\;\F{\mid}\;\D{Tm}_R\;\V{D}\;\F{\rrbracket}_R\;\D{\Uparrow}\;\V{trg}{}\<[50]%
\>[50]{}\mbox{\onelinecomment  \ensuremath{\F{active}} variables are substituted}{}\<[E]%
\ColumnHook
\end{hscode}\resethooks
It is convenient to store substitution images as a spine, because hereditary substitutions
are exactly generated from spines.
Key to the design, however, is to index \ensuremath{\D{HSub}} over the \ensuremath{\V{act}}ive subscope, as that is
what will conspicuously decrease when a recursive
substition is triggered, making \emph{termination} obvious --- one of my older
tricks~\cite{DBLP:journals/jfp/McBride03}.

Before we see how to perform a substitution, let us think how to \emph{weaken} one: we
certainly push under binders, \ensuremath{\V{jz}}, extending source and target scopes, crucially preserving
the active subscope.
\begin{hscode}\SaveRestoreHook
\column{B}{@{}>{\hspre}l<{\hspost}@{}}%
\column{3}{@{}>{\hspre}l<{\hspost}@{}}%
\column{64}{@{}>{\hspre}l<{\hspost}@{}}%
\column{E}{@{}>{\hspre}l<{\hspost}@{}}%
\>[B]{}\F{wkHSub}\;\mathbin{:}\;{}\<[64]%
\>[64]{}\D{HSub}\;\V{D}\;\V{src}\;\V{trg}\;\V{act}\;\to \;\forall\;\V{jz}\;\to \;\D{HSub}\;\V{D}\;(\V{src}\;\F{+\!\!+}\;\V{jz})\;(\V{trg}\;\F{+\!\!+}\;\V{jz})\;\V{act}{}\<[E]%
\\
\>[B]{}\F{wkHSub}\;(\V{\phi}\;\C{\sqsubseteq\![}\;\V{p}\;\C{]\!:\!=}\;\V{is})\;\V{jz}\;\mathrel{=}\;\mathkw{let}\;\C{!}\;\C{!}\;\V{p'}\;\mathrel{=}\;\F{lrCop}\;\V{jz}\;\C{[]}\;\mathkw{in}{}\<[E]%
\\
\>[B]{}\hsindent{3}{}\<[3]%
\>[3]{}(\V{\phi}\;\F{+\!\!+}_{\D{\sqsubseteq}}\;\F{oi}\;\{\mskip1.5mu \V{kz}\;\mathrel{=}\;\V{jz}\mskip1.5mu\})\;\C{\sqsubseteq\![}\;\V{p}\;\F{+\!\!+}_C\;\V{p'}\;\C{]\!:\!=}\;\F{thin\!\Uparrow}\;(\F{oi}\;\F{+\!\!+}_{\D{\sqsubseteq}}\;\F{oe}\;\{\mskip1.5mu \V{kz}\;\mathrel{=}\;\V{jz}\mskip1.5mu\})\;\V{is}{}\<[E]%
\ColumnHook
\end{hscode}\resethooks
We extend the partition to make all the bound variables passive and
duly grow the thinning on the left. On the right, co-de-Bruijn representation
lets us thin the spine of \ensuremath{\F{images}} at a stroke!

The definition of hereditary substitution is a mutual recursion, terminating because the
\ensuremath{\V{act}}ive scope is always decreasing: \ensuremath{\F{hSub}} is the main operation on terms; \ensuremath{\F{hSubs}}
proceed structurally, following a syntax description; \ensuremath{\F{hered}} handles the variable case,
invokes \ensuremath{\F{hSub}} hereditarily as required.
\begin{hscode}\SaveRestoreHook
\column{B}{@{}>{\hspre}l<{\hspost}@{}}%
\column{9}{@{}>{\hspre}l<{\hspost}@{}}%
\column{61}{@{}>{\hspre}l<{\hspost}@{}}%
\column{82}{@{}>{\hspre}l<{\hspost}@{}}%
\column{86}{@{}>{\hspre}l<{\hspost}@{}}%
\column{98}{@{}>{\hspre}l<{\hspost}@{}}%
\column{129}{@{}>{\hspre}l<{\hspost}@{}}%
\column{140}{@{}>{\hspre}l<{\hspost}@{}}%
\column{E}{@{}>{\hspre}l<{\hspost}@{}}%
\>[B]{}\F{hSub}\;{}\<[9]%
\>[9]{}\mathbin{:}\;{}\<[82]%
\>[82]{}\D{HSub}\;\V{D}\;\V{src}\;\V{trg}\;\V{act}\;\to \;\D{Tm}_R\;\V{D}\;\V{i}\;\V{iz}\;\to \;{}\<[129]%
\>[129]{}\V{iz}\;\D{\sqsubseteq}\;\V{src}\;{}\<[140]%
\>[140]{}\to \;\D{Tm}_R\;\V{D}\;\V{i}\;\D{\Uparrow}\;\V{trg}{}\<[E]%
\\
\>[B]{}\F{hSubs}\;{}\<[9]%
\>[9]{}\mathbin{:}\;{}\<[61]%
\>[61]{}(\V{S}\;\mathbin{:}\;\D{Desc}\;\V{I})\;\to \;{}\<[82]%
\>[82]{}\D{HSub}\;\V{D}\;\V{src}\;\V{trg}\;\V{act}\;\to \;{}\<[E]%
\\
\>[61]{}\F{\llbracket}\;\V{S}\;\F{\mid}\;\D{Tm}_R\;\V{D}\;\F{\rrbracket}_R\;\V{iz}\;\to \;{}\<[86]%
\>[86]{}\V{iz}\;\D{\sqsubseteq}\;\V{src}\;{}\<[98]%
\>[98]{}\to \;\F{\llbracket}\;\V{S}\;\F{\mid}\;\D{Tm}_R\;\V{D}\;\F{\rrbracket}_R\;\D{\Uparrow}\;\V{trg}{}\<[E]%
\\
\>[B]{}\F{hered}\;{}\<[9]%
\>[9]{}\mathbin{:}\;{}\<[61]%
\>[61]{}(\V{jz}\;\C{\Rightarrow}\;\V{i})\;\F{\leftarrow}\;\V{src}\;\to \;{}\<[82]%
\>[82]{}\D{HSub}\;\V{D}\;\V{src}\;\V{trg}\;\V{act}\;\to \;\F{\llbracket}\;\F{\overrightarrow{\black{\V{jz}}}}\;\F{\mid}\;\D{Tm}_R\;\V{D}\;\F{\rrbracket}_R\;\D{\Uparrow}\;\V{trg}\;\to \;\D{Tm}_R\;\V{D}\;\V{i}\;\D{\Uparrow}\;\V{trg}{}\<[E]%
\ColumnHook
\end{hscode}\resethooks

\newsavebox{\hsubs}
\savebox{\hsubs}{\hspace*{ -0.05in}
\parbox{7in}{%
\begin{hscode}\SaveRestoreHook
\column{B}{@{}>{\hspre}l<{\hspost}@{}}%
\column{25}{@{}>{\hspre}l<{\hspost}@{}}%
\column{55}{@{}>{\hspre}l<{\hspost}@{}}%
\column{59}{@{}>{\hspre}l<{\hspost}@{}}%
\column{E}{@{}>{\hspre}l<{\hspost}@{}}%
\>[B]{}\F{hSubs}\;(\C{Rec}_D\;(\V{jz}\;\C{\Rightarrow}\;\V{i}))\;{}\<[25]%
\>[25]{}\V{h}\;(\V{\theta}\;\C{\fatbslash}\;\V{t})\;{}\<[55]%
\>[55]{}\V{\psi}\;{}\<[59]%
\>[59]{}\mathrel{=}\;\V{jz}\;\F{\fatbslash}_R\;\F{hSub}\;(\F{wkHSub}\;\V{h}\;\V{jz})\;\V{t}\;(\V{\psi}\;\F{+\!\!+}_{\D{\sqsubseteq}}\;\V{\theta}){}\<[E]%
\\
\>[B]{}\F{hSubs}\;(\C{\Upsigma}_D\;\V{S}\;\V{T})\;{}\<[25]%
\>[25]{}\V{h}\;(\V{s}\;\C{,}\;\V{ts})\;{}\<[55]%
\>[55]{}\V{\psi}\;{}\<[59]%
\>[59]{}\mathrel{=}\;\F{map\!\Uparrow}\;(\V{s}\;\V{,\char95 })\;(\F{hSubs}\;(\V{T}\;\V{s})\;\V{h}\;\V{ts}\;\V{\psi}){}\<[E]%
\\
\>[B]{}\F{hSubs}\;\C{One}_D\;{}\<[25]%
\>[25]{}\V{h}\;\anonymous \;{}\<[55]%
\>[55]{}\anonymous \;{}\<[59]%
\>[59]{}\mathrel{=}\;\F{\langle\rangle}_{R}{}\<[E]%
\\
\>[B]{}\F{hSubs}\;(\V{S}\;\C{\times}_D\;\V{T})\;{}\<[25]%
\>[25]{}\V{h}\;(\C{pair}\;(\V{s}\;\C{\uparrow}\;\V{\theta})\;(\V{t}\;\C{\uparrow}\;\V{\phi})\;\anonymous )\;{}\<[55]%
\>[55]{}\V{\psi}\;{}\<[59]%
\>[59]{}\mathrel{=}\;\F{hSubs}\;\V{S}\;\V{h}\;\V{s}\;(\V{\theta}\;\F{\fatsemi}\;\V{\psi})\;\F{,}_R\;\F{hSubs}\;\V{T}\;\V{h}\;\V{t}\;(\V{\phi}\;\F{\fatsemi}\;\V{\psi}){}\<[E]%
\ColumnHook
\end{hscode}\resethooks
}
}

There is a design choice here: we may either cut the substitution down to fit the support
of the term we are processing, or retain the substitution intact and keep the thinning which
embeds the term's support in the source scope. The latter makes the termination argument
more straightforward, although we are required to curry a \ensuremath{\D{Tm}_R\;\V{D}\;\V{i}\;\D{\Uparrow}\;\V{src}} as a
\ensuremath{\V{t}\;\mathbin{:}\;\D{Tm}_R\;\V{D}\;\V{i}\;\V{iz}} with a \ensuremath{\V{\psi}\;\mathbin{:}\;\V{iz}\;\D{\sqsubseteq}\;\V{src}}. Our first move is to refine the substitution's
partition by \ensuremath{\V{\psi}} to check whether any of the variables in the term's support is actively
being substituted. If not, we may simply thin \ensuremath{\V{t}}, with no further traversal.
\begin{hscode}\SaveRestoreHook
\column{B}{@{}>{\hspre}l<{\hspost}@{}}%
\column{E}{@{}>{\hspre}l<{\hspost}@{}}%
\>[B]{}\F{hSub}\;\V{h@}\;(\V{\phi}\;\C{\sqsubseteq\![}\;\V{p'}\;\C{]\!:\!=}\;\V{is})\;\V{t}\;\V{\psi}\;\mathkw{with}\;\F{subCop}\;\V{\psi}\;\V{p'}{}\<[E]%
\\
\>[B]{}\F{hSub}\;\V{h@}\;(\V{\phi}\;\C{\sqsubseteq\![}\;\V{p'}\;\C{]\!:\!=}\;\V{is})\;\V{t}\;\V{\psi}\;\mid \;\anonymous \;\C{,}\;\C{[]}\;\C{,}\;\anonymous \;\C{,}\;\anonymous \;\C{,}\;\V{\psi_0}\;\C{,}\;\V{\psi_1}\;\C{,}\;\V{p}\;\mathkw{with}\;\F{allLeft}\;\V{p}{}\<[E]%
\\
\>[B]{}\F{hSub}\;\V{h@}\;(\V{\phi}\;\C{\sqsubseteq\![}\;\V{p'}\;\C{]\!:\!=}\;\V{is})\;\V{t}\;\V{\psi}\;\mid \;\anonymous \;\C{,}\;\C{[]}\;\C{,}\;\anonymous \;\C{,}\;\anonymous \;\C{,}\;\V{\psi_0}\;\C{,}\;\V{\psi_1}\;\C{,}\;\V{p}\;\mid \;\C{refl}\;\mathrel{=}\;\V{t}\;\C{\uparrow}\;(\V{\psi_0}\;\F{\fatsemi}\;\V{\phi}){}\<[E]%
\ColumnHook
\end{hscode}\resethooks
The \ensuremath{\C{[]}} pattern matches the active part of the support, with \ensuremath{\F{allLeft}} the lemma that
the passive part must be the whole support if the active part is empty. If, on the other
hand, there are still active variables to find, we must keep hunting, in the knowledge
that we have real work to do.
\begin{hscode}\SaveRestoreHook
\column{B}{@{}>{\hspre}l<{\hspost}@{}}%
\column{E}{@{}>{\hspre}l<{\hspost}@{}}%
\>[B]{}\F{hSub}\;\{\mskip1.5mu \V{D}\;\mathrel{=}\;\V{D}\mskip1.5mu\}\;\{\mskip1.5mu \V{i}\;\mathrel{=}\;\V{i}\mskip1.5mu\}\;\V{h@\char95 }\;\C{[}\;\V{ts}\;\C{]}\;\V{\psi}\;\mid \;\anonymous \;\mathrel{=}\;\F{map\!\Uparrow}\;\C{[}\anonymous \C{]}\;(\F{hSubs}\;(\V{D}\;\V{i})\;\V{h}\;\V{ts}\;\V{\psi}){}\<[E]%
\ColumnHook
\end{hscode}\resethooks
If we find a node from our syntax, we proceed structurally:\\
\usebox{\hsubs}

\noindent
Meanwhile, for a variable with spine attached, we substitute the spine then proceed
hereditarily.
\begin{hscode}\SaveRestoreHook
\column{B}{@{}>{\hspre}l<{\hspost}@{}}%
\column{53}{@{}>{\hspre}l<{\hspost}@{}}%
\column{E}{@{}>{\hspre}l<{\hspost}@{}}%
\>[B]{}\F{hSub}\;\V{h@\char95 }\;(\C{\scriptstyle \#}\;\{\mskip1.5mu \V{jz}\mskip1.5mu\}\;(\C{pair}\;(\C{only}\;\C{\uparrow}\;\V{x})\;(\V{ss}\;\C{\uparrow}\;\V{\theta})\;\V{c}))\;\V{\psi}\;{}\<[53]%
\>[53]{}\mid \;\anonymous \;\mathrel{=}\;\F{hered}\;(\V{x}\;\F{\fatsemi}\;\V{\psi})\;\V{h}\;(\F{hSubs}\;\F{\overrightarrow{\black{\V{jz}}}}\;\V{h}\;\V{ss}\;(\V{\theta}\;\F{\fatsemi}\;\V{\psi})){}\<[E]%
\ColumnHook
\end{hscode}\resethooks

\noindent
If the variable we seek is not the top one in the source context, we throw the top variable,
passive or active, out of the substitution and keep looking.
\begin{hscode}\SaveRestoreHook
\column{B}{@{}>{\hspre}l<{\hspost}@{}}%
\column{15}{@{}>{\hspre}l<{\hspost}@{}}%
\column{E}{@{}>{\hspre}l<{\hspost}@{}}%
\>[B]{}\F{hered}\;(\V{x}\;\C{o\apo})\;{}\<[15]%
\>[15]{}(\V{\phi}\;\C{\sqsubseteq\![}\;\V{p}\;\C{cs\apo}\;\C{]\!:\!=}\;\V{is})\;\V{ss}\;\mathrel{=}\;\F{hered}\;\V{x}\;(\F{oi}\;\C{o\apo}\;\F{\fatsemi}\;\V{\phi}\;\C{\sqsubseteq\![}\;\V{p}\;\C{]\!:\!=}\;\V{is})\;\V{ss}{}\<[E]%
\\
\>[B]{}\F{hered}\;(\V{x}\;\C{o\apo})\;{}\<[15]%
\>[15]{}(\V{\phi}\;\C{\sqsubseteq\![}\;\V{p}\;\C{c\apo s}\;\C{]\!:\!=}\;\V{is})\;\V{ss}\;\mathrel{=}\;\F{hered}\;\V{x}\;(\V{\phi}\;\C{\sqsubseteq\![}\;\V{p}\;\C{]\!:\!=}\;\F{outl}_R\;\V{is})\;\V{ss}{}\<[E]%
\ColumnHook
\end{hscode}\resethooks
We must rule out the possibility that any variable is \emph{both} active and passive.
\begin{hscode}\SaveRestoreHook
\column{B}{@{}>{\hspre}l<{\hspost}@{}}%
\column{E}{@{}>{\hspre}l<{\hspost}@{}}%
\>[B]{}\F{hered}\;\anonymous \;(\anonymous \;\C{\sqsubseteq\![}\;\anonymous \C{css}\;\{\mskip1.5mu \V{both}\;\mathrel{=}\;()\mskip1.5mu\}\;\anonymous \;\C{]\!:\!=}\;\anonymous )\;\anonymous {}\<[E]%
\ColumnHook
\end{hscode}\resethooks

\noindent
Now we have found our variable, and it is either passive (in which case we attach the spine)\ldots
\begin{hscode}\SaveRestoreHook
\column{B}{@{}>{\hspre}l<{\hspost}@{}}%
\column{15}{@{}>{\hspre}l<{\hspost}@{}}%
\column{E}{@{}>{\hspre}l<{\hspost}@{}}%
\>[B]{}\F{hered}\;(\V{x}\;\C{os})\;{}\<[15]%
\>[15]{}(\V{\phi}\;\C{\sqsubseteq\![}\;\V{p}\;\C{cs\apo}\;\C{]\!:\!=}\;\anonymous )\;\V{ss}\;\mathrel{=}\;\F{map\!\Uparrow}\;\C{\scriptstyle \#}\;(\F{va}_R\;(\F{oe}\;\C{os}\;\F{\fatsemi}\;\V{\phi})\;\F{,}_R\;\V{ss}){}\<[E]%
\ColumnHook
\end{hscode}\resethooks

\noindent
\ldots or active, in which case we substitute hereditarily.
\begin{hscode}\SaveRestoreHook
\column{B}{@{}>{\hspre}l<{\hspost}@{}}%
\column{6}{@{}>{\hspre}l<{\hspost}@{}}%
\column{9}{@{}>{\hspre}l<{\hspost}@{}}%
\column{12}{@{}>{\hspre}l<{\hspost}@{}}%
\column{15}{@{}>{\hspre}l<{\hspost}@{}}%
\column{31}{@{}>{\hspre}l<{\hspost}@{}}%
\column{49}{@{}>{\hspre}l<{\hspost}@{}}%
\column{51}{@{}>{\hspre}l<{\hspost}@{}}%
\column{E}{@{}>{\hspre}l<{\hspost}@{}}%
\>[B]{}\F{hered}\;\{\mskip1.5mu \V{trg}\;\mathrel{=}\;\V{trg}\mskip1.5mu\}\;\{\mskip1.5mu \V{act}\;\mathrel{=}\;(\anonymous \;\C{\mbox{{}-}\!,}\;(\V{jz}\;\C{\Rightarrow}\;\V{i}))\mskip1.5mu\}\;(\V{x}\;\C{os})\;{}\<[51]%
\>[51]{}(\anonymous \;\C{\sqsubseteq\![}\;\V{p}\;\C{c\apo s}\;\C{]\!:\!=}\;\V{is})\;\V{ss}{}\<[E]%
\\
\>[B]{}\hsindent{6}{}\<[6]%
\>[6]{}\mathkw{with}\;{}\<[12]%
\>[12]{}\F{outr}_R\;\V{is}\;{}\<[31]%
\>[31]{}\mid \;\F{lrCop}\;\V{trg}\;\V{jz}{}\<[E]%
\\
\>[B]{}\V{...}\;{}\<[9]%
\>[9]{}\mid \;{}\<[15]%
\>[15]{}(\V{\psi}\;\C{\fatbslash}\;\V{t})\;\C{\uparrow}\;\V{\theta}\;{}\<[31]%
\>[31]{}\mid \;\C{!}\;\C{!}\;\V{p'}\;{}\<[49]%
\>[49]{}\mathrel{=}\;\F{hSub}\;\{\mskip1.5mu \V{act}\;\mathrel{=}\;\V{jz}\mskip1.5mu\}\;(\F{oi}\;\C{\sqsubseteq\![}\;\V{p'}\;\C{]\!:\!=}\;\V{ss})\;\V{t}\;(\V{\theta}\;\F{+\!\!+}_{\D{\sqsubseteq}}\;\V{\psi}){}\<[E]%
\ColumnHook
\end{hscode}\resethooks
As you can see, the target scope becomes passive, the bound variables of the substitution
image become active, and the spine becomes the substitution for the active variables.
The new active scope is visibly a substructure of the old active scope, so hereditary
substitution is structurally recursive!

\section{Discussion}

We have a universe of syntaxes with metavariables and binding, where the \ensuremath{\D{Desc}}ription of a
syntax is interpreted as the co-de-Bruijn terms, ensuring intrinsically that unused variables are
discarded not at the \emph{latest} opportunity (as in de Bruijn terms), nor at an \emph{arbitrary}
opportunity (as in one of Bird and Paterson's variants~\cite{bird.paterson:debruijn.nested}, or
with Hendriks and van Oostrom's `adbmal' operator~\cite{DBLP:conf/cade/HendriksO03}, both of
which reduce the labour of shifting at the cost of nontrivial $\alpha$-equivalence), but at the \emph{earliest}
opportunity. Hereditary substitution exploits usage information to stop when there is nothing to substitute,
shifts without traversal, and is, moreover, structurally recursive on the \emph{active scope}.

Co-de-Bruijn representation is even less suited to human comprehension than de Bruijn syntax,
but its informative precision makes it all the more useful for machines. Dependency checking is
direct, so syntactic forms like vacuous functions or $\eta$-redexes are easy to spot.

It remains to be seen whether co-de-Bruijn representation will lead
to more efficient implementations of normalization and of metavariable
instantiation. The technique may be readily combined with representing
terms as trees whose top-level leaves are variable uses and top-level
nodes are just those (now easily detected) where paths to variables
split: edges in the tree are \emph{closed} one-hole contexts, jumped
over in constant time~\cite{conor:tube}.

I see two high-level directions emerging from this work. Firstly, the generic treatment of
syntax with \emph{meta}variables opens the way to the generic treatment of \emph{metatheory}.
Even without moving from scope-safe to type-safe term representations, we can generate
the inductive relations we use to define notions such as reduction and type synthesis
in a universe, then seek to capture good properties
(e.g., stability under substitution, leading to type soundness) by construction. Co-de-Bruijn
representations make it easy to capture properties such as variable non-occurrence in the
syntax of formul\ae, and might also serve as the target term representation
for algorithms extracted generically from the rules.

Secondly, more broadly, this work gives further evidence for a way of programming with strong
invariants and redundant but convenient information caches without fear of bugs arising
from inconsistency. We should put the programmer in charge! Dependent types should let
us take control of data representations and optimise them to support key operations,
but with the invariants clearly expressed in code and actively supporting program synthesis.

Only a fool would attempt to enforce the co-de-Bruijn invariants without support
from a typechecker, so naturally I have done so: using Haskell's {\tt
Integer} for bit vectors (making {\tt -1} the identity of the unscoped
thinning \emph{monoid}), I implemented a dependent type system, just for
fun. It was Hell's delight, even with the Agda version to follow. I was
sixteen again.

\paragraph{Acknowledgements.~} EPSRC project
EP/M016951/1 \emph{Homotopy Type Theory: Programming and Verification} funded
this work. My Mathematically Structured
Programming colleagues at Strathclyde made me get these ideas in shape:
Fredrik Nordvall Forsberg offered particularly useful advice about what
to omit. Philippa Cowderoy's use of \emph{information effects} for
typing contexts increased my sensitivity to the signposting of discards and duplications.
An EU TYPES Short Term Scientific Mission brought Andrea Vezzosi
to Strathclyde, provoking ideas and action for further work.
Invitations to present at \emph{Trends in Functional Programming 2017}
(Canterbury) and in Nottingham (with Thorsten Altenkirch in the audience) helped
me find the words.
Andreas Abel sent some particularly helpful feedback, as did a number of
commenters on social media. The advice of anonymous referees has instigated
significant improvement.
As ever, James McKinna and Randy Pollack remain sources of inspiration.

\bibliographystyle{eptcs}
\bibliography{EGTBS}
\end{document}